\documentclass{aastex63}

\usepackage{color}
\usepackage{natbib}
\usepackage{graphicx}
\usepackage{hyperref}
\usepackage{amssymb}
\usepackage{multirow}
\usepackage{chngcntr}
\usepackage[varg]{txfonts}

\def\baselinestretch{1}
\bibpunct{(}{)}{;}{a}{}{,}

\hypersetup{
    bookmarks=true,         % show bookmarks bar?
    unicode=false,          % non-Latin characters in Acrobat’s bookmarks
    pdftoolbar=true,        % show Acrobat’s toolbar?
    pdfmenubar=true,        % show Acrobat’s menu?
    pdffitwindow=true,     % window fit to page when opened
    pdfstartview={FitH},    % fits the width of the page to the window
    pdftitle={},    % title
    pdfauthor={Beers},     % author
    pdfsubject={Astronomy},   % subject of the document
    pdfcreator={dvipdf},   % creator of the document
    pdfproducer={dvipdf}, % producer of the document
    pdfkeywords={metal-poor stars},% list of keywords
    pdfnewwindow=true,      % links in new window
    colorlinks=true,       % false: boxed links; true: colored links
    linkcolor=red,          % color of internal links
    citecolor=blue,        % color of links to bibliography
    filecolor=magenta,      % color of file links
    urlcolor=cyan,           % color of external links
    breaklinks=true,
    linktocpage
}

\begin{document}
\title{Evidence for the Third Stellar Population in the Milky Way's Disk} 

\author{Daniela Carollo}
\affiliation{INAF, Astrophysical Observatory of Turin, Torino, Italy}
\author{Masashi Chiba}
\affiliation{Astronomical Institute, Tohoku University, Sendai 980-8578, Japan}
\author{Miho Ishigaki}
\affiliation{Astronomical Institute, Tohoku University, Sendai 980-8578, Japan}
\author{Ken Freeman}
\affiliation{ANU - Research School of Astronomy and Astrophysics, Weston - ACT, Australia}
\author{Timothy C. Beers}
\affiliation{Department of Physics and JINA Center for the Evolution of the Elements, University of Notre Dame, Notre Dame, IN  46556, USA}
\author{Young Sun Lee}
\affiliation{Department of Astronomy and Space Science, Chungnam National University,Daejeon 34134, Korea}
\author{Patricia Tissera}
\affiliation{Departamento de Ciencias Fisicas, Universidad Andres Bello, Av. Republica 220, Santiago, Chile}
\author{Chiara Battistini}
\affiliation{Zentrum f\"ur Astronomie der Universitat Heidelberg, Landessternwarte, Germany}
\author{Francesca Primas}
\affiliation{ European Southern Observatory, Schwarzschild-Str. 2, 85748 Garching, Germany}

%\affiliation{\altaffilmark{1} INAF, Astrophysical Observatory of Turin, Torino, Italy}
%\affiliation{\altaffilmark{2} Astronomical Institute, Tohoku University, Sendai %980-8578, Japan}
%\affiliation{\altaffilmark{3} ANU - Research School of Astronomy and %Astrophysics, Weston - ACT, Australia}
%\affiliation{\altaffilmark{4} Department of Physics and JINA Center for the %Evolution of the Elements,
% University of Notre Dame, Notre Dame, IN  46556, USA}
%\affiliation{\altaffilmark{5} Department of Astronomy and Space Science, %Chungnam National University,Daejeon 34134, Korea}
%\affiliation{\altaffilmark{6} Departamento de Ciencias Fisicas, Universidad Andres %Bello, Av. Republica 220, Santiago, Chile}
%\affiliation{\altaffilmark{7} Zentrum f\"ur Astronomie der Universitat Heidelberg, %Landessternwarte, Germany}
%\affiliation{\altaffilmark{8} European Southern Observatory, Schwarzschild-Str. 2, %85748 Garching, Germany}

\begin{abstract} 
The Milky Way is a unique laboratory, where stellar properties can be measured and 
analyzed in detail. In particular, stars in the older populations encode information 
on the mechanisms that led to the formation of our Galaxy. In this article, we analyze 
the kinematics, spatial distribution, and chemistry of a large number of stars in 
the Solar Neighborhood, where all of the main Galactic components are 
well-represented. We find that the thick disk comprises two distinct and overlapping 
stellar populations, with different kinematic properties and chemical compositions. 
The metal-weak thick disk (MWTD) contains two times less metal content than the 
canonical thick disk, and exhibits enrichment of light elements typical of the oldest 
stellar populations of the Galaxy. The rotational velocity of the MWTD around the 
Galactic center is $\sim$ 150 km s$^{-1}$, corresponding to a rotational lag of 30 
km s$^{-1}$ relative to the canonical thick disk ($\sim$ 180 km s$^{-1}$), with a 
velocity dispersion of 60 km s$^{-1}$. This stellar population likely originated from 
the merger of a dwarf galaxy during the early phases of our Galaxy's assembly, or 
it is a precursor disk, formed in the inner Galaxy and brought into the Solar 
Neighborhood by bar instability or spiral-arm formation mechanisms.   
\end{abstract}


\section{Introduction}

One of the most actively explored areas of contemporary astronomy is the formation 
and evolution of Milky Way-like galaxies and their main structural components.  It 
is now possible to perform accurate chemo-dynamical studies of stellar populations 
in our Galaxy, thanks to the advent of large-scale surveys such as the Sloan Digital 
Sky Survey \citep[SDSS;][]{york2000}, LAMOST \citep{cui2012}, and 
the Gaia mission \citep{brown2018}. The 
photometry and medium-resolution spectroscopy from SDSS or LAMOST provides estimates of stellar atmospheric parameters (effective temperature, surface gravity, and 
metallicity), and radial velocities, while the Gaia satellite delivers accurate 
positions, trigonometric parallaxes, and proper motions for many millions of stars. 
The combination of chemical information and kinematics is a powerful tool to explore 
the structures of the Milky Way, and to derive the properties of the associated stellar 
populations, defined as samples of stars that exhibit common spatial distributions, 
kinematics, and chemical composition.\par
In the case of old stellar populations in the Milky Way (the bulge, thick disk, and 
halo system), spatial motions and chemical abundances can reveal crucial information 
on the assembly history of the Galaxy, because these stars conserve the kinematic 
and chemical properties of the ancient Galactic building blocks and assembly 
mechanisms \citep{bullock2005,tissera2010,moore2006,brook2007}. 
Previous works have successfully employed kinematics and chemical abundances to 
analyze the complexity of the Galaxy's halo, and have revealed the presence of (at 
least) two diffuse stellar components \citep{carollo2007,carollo2010, beers2012, tian2019} and numerous structures 
and over-densities \citep{grillmair2016}.\\  Recent works based on the second Gaia data release \citep[DR2;][]{brown2018} have shown the presence of additional structures or debris in the halo system, likely relics of past merging events \citep{helmi2018, myeong2019}.\par
The thick disk (hereafter TD), one of the main structural components of the Milky Way, has been 
extensively studied since its discovery \citep{yoshii1982,gilmore1983}. 
Stars in the TD possess different kinematics and chemical composition from the younger 
thin-disk stellar population. The metallicity distribution of the TD exhibits a peak 
at [Fe/H] = $-$0.6 \citep{chiba2000,carollo2010,ivezic2008,lee2011b},
however, stars with disk-like kinematics have been observed at even 
lower metallicity, down to [Fe/H] = $-$1.7 \citep{chiba2000,carollo2010}, and possibly lower. 
This low-metallicity tail of the TD is known as the metal-weak thick disk
\citep[MWTD;][]{morrison1990,beers2014}.\par
The properties of the MWTD, such as its velocity components, metallicity range, and 
stellar orbital parameters have been challenging to derive, due to the strong overlap 
with the TD and inner stellar halo, as well as the lack of accurate stellar distances and 
intrinsic motions. Previous attempts to model the MWTD as an independent stellar 
population from the TD \citep{carollo2010,ivezic2008} have revealed that 
its mean Galactocentric rotational velocity could be in the range 100 - 150 km s$^{-1}$, 
while its metallicity spans values from [Fe/H] =  $-$0.8 to $-$1.7.\par
In this article, we explore the nature of the MWTD and present evidence that it is 
an {\it independent} stellar population,  using a sample of SDSS stars combined with  
accurate astrometric parameters provided by the second Gaia data release.

\section{Methods}

The stellar disk and halo populations are defined by their fundamental properties 
in kinematics and chemical space. For the stellar halo, the quantity $V_{\rm rot}/\sigma$
(where $V_{\rm rot}$  is the mean rotational velocity around the 
Galaxy center, and $\sigma$ is the total velocity dispersion) is smaller than 1, 
indicating that halo stars have random motions around the Galactic center, with high 
energy, and can achieve large distances during their orbits. Also, their kinematic 
parameters and the spatial density do not exhibit strong dependence on the Galactocentric 
distance ($R$) or distance from the Galaxy's plane ($|Z|$), in the Solar Neighborhood. 
On the contrary, the disk components exhibit a ratio $V_{\rm rot}/\sigma$ larger 
than 1, suggesting a large mean rotation, and their kinematic and spatial density 
depend on the $R$ and $Z$ distances. Galaxy disks show a highly flattened distribution 
along the $|$Z$|$ direction as well. In chemical space, the disk and halo components 
differ in the content of metals ([Fe/H]) and their $\alpha$-element ratios 
([$\alpha$/Fe]). Halo stars are metal-poor ([Fe/H] $< -1.0$) and $\alpha$-element 
rich. The two previously recognized disk components, the thin disk and the TD, have 
different chemical properties $-$ the thin disk is metal-rich 
($\langle$[Fe/H]$\rangle \sim 0$), and exhibits a low 
 $\langle$[$\alpha$/Fe]$\rangle$ ratio, while the TD is more metal-poor 
($\langle$[Fe/H]$\rangle \sim -0.6$), and exhibits a higher [$\alpha$/Fe] ratio. 
These definitions and properties should be considered when performing kinematics and 
chemical analysis of Galactic stellar populations to avoid a mixture of definitions. 

\subsection{Selection of the Data}

In this work, we adopt a sample of SDSS calibration stars observed during the course 
of the SEGUE (Sloan Extension for Galactic Understanding and
Exploration) sub-survey \citep{yanny2009}, carried out as part of SDSS-II from 2005 to 2008. The 
SEGUE survey obtained $\sim$ 240,000 medium-resolution spectra of stars in the 
Galaxy, with the aim to study the stellar populations from 0.5 kpc to 100 kpc. 
SDSS used a 2.5 m telescope located at Apache Point Observatory in New Mexico, and 
equipped with an imaging camera and two spectrographs, capable of observing 640 
spectra in total, over a 7 deg$^{\rm 2}$ field of view. For each spectroscopic 
plug-plate, spectroscopy was obtained for a small number of stars (16) selected to 
remove the distortions of the observed flux of stars and galaxies arising from the 
wavelength response of the ARC 2.5m telescope and the SDSS spectrographs, as well 
as the distortions imposed on the observed spectra by the Earth's atmosphere. This 
set of calibration stars has apparent magnitude in the range 15.5 $<$ g$_{0}$ $<$ 17.0, 
and color ranges 0.6 $<$ (u $-$ g)$_{0}$ $<$ 1.2 and  0 $<$ (g $-$ r)$_{0}$ $<$ 0.6. 
The subscript 0 in the magnitudes and colors indicates that they are corrected for 
the effects of interstellar absorption and reddening, following standard procedures 
\citep{schlegel1998}.  The signal-to-noise for these stars' spectra is typically 
S/N $>$ 20. A second set of calibration stars was used to calibrate and remove from 
SDSS spectra the presence of night-sky emission and absorption features (telluric 
calibration stars). Such stars have the same color range as the first set, but at 
fainter apparent magnitudes, in the range 17.0 $<$  g$_{0} <$ 18.5, and 
signal-to-noise of 20 $<$ S/N $<$ 30. The majority of the 
flux- and telluric-calibration stars are located close 
to the main-sequence turnoff.\par
Stellar parameters are obtained using the SEGUE Stellar Parameter Pipeline
\citep[SSPP;][]{lee2008a,lee2008b, lee2011a},
which processes the wavelength- and 
flux-calibrated spectra generated by the standard SDSS spectroscopic reduction 
pipeline, obtains equivalent widths and/or line indices for about 80 atomic or 
molecular absorption lines, and estimates the effective temperature, $T_{\rm eff}$, surface 
gravity, $\log g$, and metallicity, [Fe/H], for a given star through the application 
of a number of approaches. A given method is usually optimal over specific ranges 
of color and signal-to-noise ratio \citep{lee2008a}. Typical internal errors for 
stars in the temperature range that applies to the calibration stars are 
$\sigma(T_{\rm eff}) \sim$ 100 K to 125 K, $\sigma(\log g) \sim$ 0.25 dex, and 
$\sigma$ ([Fe/H]) $\sim$ 0.20 dex.  The external errors in these determinations are 
of similar size.\par
The $\alpha$-element abundances are also derived by the SSPP. The pipeline makes use 
of a pre-existing grid of synthetic spectra \citep[NEWODF;][]{castelli2003}, with no 
enhancement in $\alpha$-element abundances, and creates a fine (steps of 0.2 dex for 
$\log g$ and 0.2 dex for [Fe/H]) grid of spectra by interpolation between the wider model 
grids (steps of 0.5 dex). The wavelength range is 4500 $-$ 5500 $\AA$, chosen because 
it contains a large set of metallic lines, but avoids the CH $G$-band feature 
($\sim$ 4300 $\AA$, which can be strong in metal-poor stars) and the Ca II K 
( $\sim$ 3933 \AA) and H ($\sim$ 3968 $\AA$) lines, which can saturate for cool 
metal-rich stars. The final grid covers 4000 K $<$ $T_{\rm eff}$  $<$ 8000 K, in steps 
of 250 K, 0.0 $<$ $\log g$ $<$ 5.0, in steps of 0.2 dex, and $-$4.0 $<$ [Fe/H] $< -$0.4, 
in steps of 0.2 dex. The range in [$\alpha$/Fe] introduced for the spectral synthesis 
covers +0.1 $<$ [$\alpha$/Fe] $<$ +0.6, in steps of 0.1 dex, at each node of $T_{\rm eff}$, 
$\log g$, and [Fe/H]. After creation of the full set of synthetic spectra, they are 
degraded to SEGUE resolution (R = 2000) and re-sampled to 1 $\AA$ wide linear pixels 
(during SSPP processing, the SEGUE spectra are also linearly re-binned to 1 $\AA$ per 
pixel). In the SSPP pipeline, the notation [$\alpha$/Fe] denotes an average of the 
abundance ratios for individual $\alpha$-elements weighted by their line strengths 
in synthesized spectra. In the selected spectral range the dominant features are the 
magnesium (Mg) and titanium (Ti) lines, which are the primary contributors to the 
determination of [$\alpha$/Fe], with some influence from silicon (Si) and calcium 
(Ca). In the adopted wavelength range, and at the SDSS spectral resolution, oxygen 
(O) has no strong detectable features, and it is excluded in the computation of the 
overall $\alpha$-element abundance. The [$\alpha$/Fe] measurements were validated 
with the stars in other external sources such as the ELODIE \citep{prugniel2001} 
spectral library, and compared with those obtained by analyzing a large sample of 
SEGUE stars observed at high spectral resolution. The SSPP provides [$\alpha$/Fe] 
abundance for SDSS/SEGUE spectra with a precision of $\sim$ 0.06 dex at S/N $>$ 50 
and $<$ 0.1 dex at S/N = 20.\par
Radial velocities for stars in our sample are derived from matches to an external 
library of high-resolution spectral templates with accurately known velocities 
\citep{Allende2007,yanny2009}, degraded in resolution to that of the SDSS 
spectra. The typical precision of the resulting radial velocities are on the order 
of 3-20 km s$^{-1}$, depending on the S/N of the spectra, with zero-point errors of no 
more than 3 km s$^{-1}$, based on a comparison of the subset of stars in our sample 
with radial velocities obtained from the high-resolution spectra taken for testing 
and validation of the SSPP. \par
The initial sample employed in this analysis consists of $\sim$ 32,000 unique stars 
selected in the temperature range 4500 K $<$ $T_{\rm eff}$ $<$ 7000 K, where the SSPP 
pipeline provides the highest accuracy for the derived atmospheric parameters. This 
sample is cross-matched with the Gaia DR2 database to retrieve accurate astrometric 
positions, trigonometric parallaxes, and proper motions, using the CDS (Centre de 
Données Astronomiques de Strasbourg) X-Match service, adopting a very small search 
radius (0.8'') to avoid duplicates. The match provides positions, parallaxes, and 
proper motions for all of the stars in the sample. We select stars with relative 
parallax errors of $\sigma_{\pi}/\pi <$ 0.2, and derive their distance estimates 
using the relation $d = 1/\pi$ (and used these distances to select stars with 
heliocentric distance d $\leq$ 4 kpc). This selection reduces the number of stars 
to 10,820 ($\sim$ 1/3rd of the initial sample). The majority of stars in this final 
sample have errors on proper motions below 0.2 mas yr$^{-1}$. We also select stars 
with Galactocentric distance in the range 7 kpc $<$ $R$ $<$ 10 kpc which reduces the 
number of stars to 9,258.\par
Regarding the systematic error on the parallaxes, the so-called parallax zero-point 
offset \citep{luri2018}, most of the stars have parallax  $\pi >$ 0.2 mas (only four 
stars have parallax below this value), and our distances should not be affected 
significantly from such an offset. Considerations on possible effects of biases on the parallax are discussed in Appendix A1, where it is shown that the addition of the parallax zero-point offset has only a minor effect on the results presented in this article. 
\clearpage

\subsection{Derivation of Kinematic and Dynamic Parameters}

The proper motions, in combination with distances and radial velocities, provide the 
information required to calculate the full space motions. The components of the space 
motions are represented as the ($U,V,W$) velocities of the stars with respect to the 
Local Standard of Rest (LSR).  The velocity component $U$ is taken to be positive in the direction toward the
Galactic anticenter, the V component is positive in the direction of Galactic
rotation, and the W component is positive toward the North Galactic Pole. Corrections for the motion of 
the Sun with respect to the LSR are applied during the course of the calculation of 
the full space motions; here we adopt the values ($U, V, W$) = ($-$9, 12, 7) km 
s$^{-1}$ \citep{mihalas1981}. We also obtained the rotational component of a star's 
motion around the Galactic center in a cylindrical frame; this is denoted as 
$V_{\phi}$, and is calculated assuming that the LSR is on a circular orbit with a 
value of 220 km s$^{-1}$ \citep{bovy2012a}. The orbital parameters $(r_{\rm apo},
r_{\rm peri}, Z_{\rm max}, {\rm eccentricity})$ and integrals of motions
are derived by adopting an analytic St\"ackel-type gravitational potential (which consists of 
a flattened, oblate disk and a nearly spherical massive halo), and integrating their 
orbital paths based on the starting point obtained from the observations 
\citep{chiba2000}. \par

We derived the vertical angular momentum component, $L_{\rm Z}  = 
V_{\phi} R$ (where $V_{\phi}$ is the Galactocentric rotational velocity and $R$ 
is the distance from the Galaxy's center projected on to its plane), and the perpendicular 
angular momentum, $L_{\rm P} = (L_{\rm X}^2 + L_{\rm Y}^2)^{1/2}$
 (where $L_{\rm X}$ and $L_{\rm Y}$ represent orthogonal angular 
momentum components in the plane of the Galaxy). They are defined as $L_{\rm X} = Y V_{\rm Z} - Z V_{\rm Y}$ 
and $L_{\rm Y} = Z V_{\rm X} - X V_{\rm Z}$, $(X,Y,Z)$ is the 
star's position in the Galactocentric Cartesian reference frame, and $(V_{\rm X},
V_{\rm Y},V_{\rm Z})$ are the velocity components in this system. These parameters are 
obtained by adopting for the Sun's location, $R_{\rm Sun}$ = 8.5 kpc \citep{ghez2008}, 
and a circular velocity at the position of the Sun of 220 km s$^{-1}$ \citep{bovy2012a}. \par

\section{Results: Evidence for a second stellar population in the thick disk}

Figure 1 shows the distribution of stars (with relative errors on parallax less than 
20\%, and  heliocentric distances less than 4 kpc), in the plane defined by 
L$_{\rm Z}$ and $L_{\rm P}$. Stars that rotate in the same direction as the 
Galactic rotation (prograde) have positive values of L$_{\rm Z}$. 
In this diagram, the stars that come from the same 
progenitor(s) might be expected to cluster together, as they have similar values in 
the (L$_{\rm Z}$, L$_{\rm P})$ plane.\par
The full sample of stars shown in Figure 1a exhibits two groups, with prograde rotation 
at $L_{\rm Z}$ $\sim$ 1800 kpc km s$^{-1}$ and $L_{\rm Z}$ $\sim$ 1200 kpc km s$^{-1}$, 
respectively. The group with larger vertical angular momentum, corresponding to a 
larger Galactocentric rotational velocity, represents the superposition of the 
thin-disk and TD components \citep{carollo2010}. The less-prograde group of stars 
has rotational properties not previously associated with a separate stellar 
population in the Milky Way -- this is the MWTD, which, up to now, appeared to be 
consistent with the low-metallicity tail of the TD, but as we demonstrate below, is 
a distinct population. \par
Figure 1b represents the metallicity, as a function of $L_{\rm Z}$, for the entire 
stellar sample, color-coded on the relative abundance ratio between the 
${\alpha}$-elements (e.g., Mg, Ti) and metallicity ([${\alpha}$/Fe]). Typical values 
of [${\alpha}$/Fe] for the thin disk are between 0.0 and +0.15 \citep{lee2011b,haywood2013,
recioblanco2014,hayden2015}, with metallicities 
$-$0.3 $<$ [Fe/H] $<$ +0.5. However, some investigations report evidence of the 
existence of a sub-population of thin-disk stars with metallicities $-$0.6 $<$ [Fe/H] 
$<$ $-$0.2 and 0.0 $<$ [${\alpha}$/Fe] $<$ +0.2, located at Galactocentric distance 
$R$ $>$ 9 kpc, and possessing large mean rotational velocity \citep{lee2011b,  
bovy2012b}. These lower metallicity outer thin-disk stars are partially 
present in our sample (we estimate a fraction of $\sim$ 14\% among the stars at $R$ 
$>$ 9 kpc; see Section 4), however they don't affect the properties of the 
less-prograde stellar population, and for the rest of the analysis we only consider 
their influence on the parameters derived for the TD.  
In Figure 1b, stars with +0.15 $<$ [${\alpha}$/Fe] $<$ +0.2 are represented by a cyan 
color, while stars with [${\alpha}$/Fe] $>$ +0.22 are shown in red. From inspection 
of this figure, the  more prograde stellar population comprising the TD and some 
thin-disk contamination, and the MWTD-like populations separate very well in the 
($L_{\rm Z}$, [Fe/H]) plane. The rest of the stars, with large values of 
[${\alpha}$/Fe] and lower metallicity, belong to the halo system.
The average error on $L_{\rm Z}$ and $L_{\rm P}$ is 100 kpc km s$^{-1}$ and 70 kpc km s$^{-1}$, respectively; they are represented by error bars legend in Figure 1a.\par

The peaks seen in Figure 1c clearly show the separation of the more-prograde and the 
less-prograde stellar components in the vertical angular momentum distribution. 
Rotational properties similar to those exhibited by the MWTD-like component were 
obtained in a previous work, assuming that the MWTD could be treated as an independent 
stellar population (100 km s$^{-1} <$ $V_{\phi} <$ 150 km s$^{-1}$,  
\citealt{carollo2010}). Average rotational velocities for the thin disk and TD are 220 
km s$^{-1}$ \citep{rix2013} and 180 km s$^{-1}$ \citep{carollo2010}, respectively.\par
Figure 1d shows the normalized density distribution for the full sample (gray), the 
more-prograde  stellar populations (cyan) and the high-${\alpha}$ components (red), 
which are dominated by the MWTD-like and halo populations (the TD is also present 
in this range of $\alpha$-element abundances). It is worth noting that the peak in 
$L_{\rm Z}$ for the cyan cluster ($L_{\rm Z} \sim$ 1800-2000 kpc km s$^{-1}$) is larger 
than the known average value for the TD ($L_{\rm Z} \sim$ 1600 kpc km s$^{-1}$, 
assuming the Sun's location at 8.5 kpc), due to  thin-disk contamination.  
The panels of Figure 1 show, for the first time, clear evidence that the MWTD-like stellar 
population has different kinematics and chemical composition than the TD and halo, suggesting a distinct 
astrophysical origin.\par

Figure 2 shows [${\alpha}$/Fe], as a function of the metallicity, for the full sample 
of stars (panel a), and a sub-sample obtained by selecting stars with vertical angular 
momentum $L_{\rm Z} >$ 500 kpc km s$^{-1}$ and distance from the plane  $|Z|$ $>$ 1 
kpc (panel b). The cut in angular momentum reduces the contamination from halo stars, 
while the cut in vertical distance removes most of the thin-disk stars.\par

In Figure 2a, stars with metallicity $-$0.9 $<$ [Fe/H] $<$ $-$0.2 and +0.1 
$<$ [${\alpha}$/Fe] $<$ +0.25 represent the abundance sequence of the TD 
\citep{hayden2015} with some thin-disk contamination, including the outer thin-disk 
stars that can reach large distances from the Galactic plane \citep{lee2011b, 
bovy2012b,savino2019}. The sequence delimited by the abundance 
ranges $-$1.2 $<$ [Fe/H] $<$ $-$0.6 and +0.22 $<$ [${\alpha}$/Fe] $<$ +0.35 is 
dominated by the prograde component with lower vertical angular momentum identified 
in Figure 1 (MWTD-like). The rest of the distribution, with a range in abundances 
[Fe/H] $<$  $-$1.2 and  +0.2 $<$ [${\alpha}$/Fe]  $<$ +0.32, belongs to the halo 
system of the Galaxy.\par
The  MWTD-like sequence is well-represented in the sub-sample shown in Figure 2b, 
although some stars in this stellar population were removed by the adopted cuts in 
$L_{\rm Z}$ and [Fe/H]. It is worth noting that in Figure 2b, the two sequences in the 
range $-$1.2 $<$ [Fe/H] $<$ $-$0.3 are neither pure TD or pure MWTD.  The TD (including 
outer thin-disk contaminants) and the MWTD-like stellar populations are strongly 
overlapped in the abundance space, and they dominate in the range of metallicity 
$-$1.2 $<$ [Fe/H] $<$ $-$0.3 with ${\alpha}$-element ratios +0.1 $<$ [${\alpha}$/Fe]  
$<$ +0.37. \par
Figure 2c shows the metallicity distribution for the same sample of stars as in Figure 
2b, but divided into two sub-samples, the ${\alpha}$-poor ([${\alpha}$/Fe]  
$<$ +0.22) and ${\alpha}$-rich ([${\alpha}$/Fe]  $>$ +0.22) stars. This distribution 
exhibits two metallicity peaks, at  [Fe/H] $\sim$ $-$0.6 (cyan; right edge of the maxima of the distribution) and $-$1.0  (red; left edge of the maxima of the distribution), that correspond 
to the TD and MWTD-like stellar populations, with some contamination from lower 
metallicity halo stars ([Fe/H] $\sim$ $-$1.6) and outer thin-disk stars, which are identified 
in the abundance space with [Fe/H] up to $-$0.7 and +0.1 $<$ [${\alpha}$/Fe] $<$ +0.2 
\citep{lee2011b, bovy2012b}. Figure 2d shows the [${\alpha}$/Fe] abundance 
distribution for the sample of stars from Figure 2b, with an additional cut in 
metallicity to minimize the halo-star contamination. A mixture model analysis employing two independent Gaussian components applied to this distribution provides a very good fit, with mean values of [${\alpha}$/Fe] = $+$0.18 and [${\alpha}$/Fe] = $+$0.28, and  standard deviations $\sigma_{1}$ = 0.035 dex, and $\sigma_{2}$ = 0.036 dex, respectively. The low-${\alpha}$ component (TD dominated) accounts for 34\% of the total number of stars in the sub-sample , while the high-${\alpha}$ component (MWTD dominated) accounts for 66\%.  In Figure 2d, the low-${\alpha}$  and high-${\alpha}$ Gaussian fits are displayed in cyan and red colors, respectively.\par
Figure 3 shows the distribution of the logarithmic number density, over-plotted with 
equidensity contours for the entire sample in the ($L_{\rm Z}, L_{\rm P}$) and 
([Fe/H], $L_{\rm Z}$) planes (panel a and b, respectively). The two prograde 
components can be clearly identified with the high-density reddish clusters with 
equidensity contour of 1.2 (panel a) and 1.0 (panel b). In panels c and d, the density 
distributions are represented in the ([${\alpha}$/Fe], [Fe/H]) plane for the entire 
sample, and for a sub-sample obtained by selecting stars with vertical angular 
momentum $L_{\rm Z}$ $>$ 500 kpc km s$^{-1}$ and distance from the plane of $|Z| 
>$ 1 kpc, respectively, as in Figure 2. The density contour plots provide better 
visualization of the various stellar populations described in Figure 1 and 2.\par
The rotational properties and abundances are key parameters for the separation of 
overlapping stellar populations in the Milky Way \citep{chiba2000, 
carollo2007, carollo2010}. For the remaining analysis, we adopted a sub-sample of stars 
with metallicity $-$1.2 $<$ [Fe/H] $<$ $-$0.6 and ${\alpha}$-element abundances 
[${\alpha}$/Fe] $>$ +0.2 (Sample A).  These cuts select stars belonging mainly to 
the MWTD-like and TD components, with some contamination from the inner halo (IH), 
while the previously noted contamination from the outer thin-disk stars is not present in this range 
of metallicity and ${\alpha}$-element abundances.\par
Figure 4 shows the distribution of the vertical angular momentum (left column of 
panels) and (signed) orbital eccentricity (right column of panels) of Sample A for 
different cuts in the absolute values of vertical distance, $|Z|$, where the plus 
(minus) signs in eccentricity denote prograde (retrograde) motions. As a comparison, 
an additional sub-sample (Sample B), comprising mainly TD stars, obtained by adopting 
the cuts $-$0.9 $<$ [Fe/H] $<$ $-$0.6 and +0.1 $<$ [${\alpha}$/Fe] $<$ +0.2, is 
over-plotted in red. This sample also has some thin-disk contamination. \par
Visual inspection of the Sample A distributions reveals that, up to a vertical 
distance of $|Z|$ = 1 kpc, the TD and MWTD-like stellar populations are strongly 
overlapped, with some contamination from the IH, while Sample B is completely 
dominated by the overlapping thin disk and TD ($L_{\rm Z} \sim$ 2000 kpc km 
s$^{-1}$). As expected, close to the Galactic plane, most of the stars have low orbital 
eccentricities (ecc $\sim$ +0.1). At larger distances, 1 kpc $<$ $|Z|$ $<$ 2 kpc, 
the dominant components in Sample A are the MWTD-like and TD populations, with some 
contribution from the IH ($L_{\rm Z} \sim$ 1200 kpc km s$^{-1}$,
$L_{\rm Z} \sim$ 1800 kpc km s$^{-1}$, and $L_{\rm Z} \sim$ 0 kpc km s$^{-1}$, 
respectively).\par
The orbital eccentricity distributions shown in the right column of panels has stars 
with larger values associated with the MWTD-like component (up to ecc $\sim$ +0.5) 
and the IH \citep[$|$ecc$| >$ 0.5;][]{carollo2010}. In this range of distances, Sample 
B is still dominated by the overlapping thin disk and TD, with $L_{\rm Z} \sim$ 1800 
kpc km s$^{-1}$ and ecc $\sim$ +0.1.   In the range 2 kpc $<$ $|Z| <$ 3 kpc, the 
MWTD is the dominant component in Sample A, with a larger contribution from the IH 
($L_{\rm Z} \sim$ 1200 kpc km s$^{-1}$ and $L_{\rm Z} \sim$ 0 kpc km s$^{-1}$), 
while Sample B is TD-dominated, although the number of stars at these vertical 
distances is rather low (the histograms are normalized to 1 to allow a comparison 
between the two samples).  
This is also evident in the orbital eccentricity distribution, where most of the stars 
possess +0.2 $<$ ecc $<$ +0.6 (and ecc $<$ $-$0.5, corresponding to IH stars). At 
distances $|Z|$ $>$  3 kpc, the MWTD-like stellar population is still present, but 
it is very weak. In all the panels the distributions are normalized to facilitate a visual comparison between samples, in particular, in those cases where Sample B contains  a much lower number of stars than Sample A (at larger vertical distances). The peaks of the $L_{\rm Z}$ distributions include 95, 308, 92, and 6 stars, respectively, for Sample A, and 60, 88, 13, 1 stars, respectively, for Sample B. \par
A mixture model analysis employing three independent (Gaussian) components applied 
to the distributions of stars in Sample A, up to $|$Z$|$ = 3 kpc, provides estimates 
for the mean vertical angular momentum and dispersion reported in Table 1. The mixture model includes three independent components identified in the table as TD, MWTD-like, and IH. Close to the Galactic plane, the TD and MWTD dominate the distribution, with 43\% and 38\% stellar fractions. At larger vertical distance the main component (60\% of the stars)  is MWTD-like, with a mean angular momentum of $\sim$ 1100 to $\sim$ 1200  kpc km s$^{-1}$. The inner-halo stellar population accounts for 19\%, 14\% and 30\% of the total number of stars in Sample A, respectively, while the TD-dominated component contributes 43\%, 26\%, and 8\%, at 0 kpc $< |$Z$| <$ 1 kpc, 1 kpc $< |$Z$| <$ 2 kpc, and 2 kpc $< |$Z$| <$ 3 kpc, respectively.\par
It is well-known that the thick disk exhibits a negative gradient in the mean rotational velocity as a function of the vertical distance (\citet*{chiba2000}; \citet*{carollo2010}, and references therein). This is an intrinsic property of this stellar population, which is likely due to its mechanism of formation. By using a sub-sample of stars in the range $-$1.1 $<$ [Fe/H] $<$  $-$0.6 and [${\alpha}$/Fe] $>$ +0.22, we show that the MWTD-like population exhibits a dependence on the vertical distance (Figure 5), with a gradient of 
$\Delta \langle V_{\phi} \rangle/\Delta \langle |Z| \rangle = -21$ km 
s$^{-1}$ kpc$^{-1}$, lower than that determined for the canonical TD 
($\Delta \langle V_{\phi} \rangle/\Delta \langle |Z| \rangle = -36$ km 
s$^{-1}$ kpc$^{-1}$, \citealt{carollo2010}). 
Determination of the mean rotational 
velocity and its dispersion for the MWTD is accomplished by assuming a fiducial sample of stars 
selected in the range 1 kpc $<$ $|Z|$ $<$ 2 kpc, $-$1.1 $<$ [Fe/H] $<$ $-$0.6, and 
[${\alpha}$/Fe] $>$ +0.25. The results are $\langle V_{\phi} \rangle$ = 147 $\pm$ 2 
km s$^{-1}$ and $\sigma_{V_{\phi}}$ = 60 $\pm$ 2 km s$^{-1}$. Note that a similar 
velocity dispersion was obtained in a recent work \citep{robin2017} for the old thick-disk 
stellar population, corresponding to the oldest episode of star formation in 
the disk (12 Gyr), according to the Besancon population synthesis model  
\citep{robin2014}. \citet{tian2019}
idenfity a disk-like component with mean rotation and velocity dispersion which
they claimed could be identified with the MWTD, but it appears that
their values are more consistent with the canonical TD.\par
The mean orbital parameters, such as the apo-Galactic and peri-Galactic distances 
($\langle r_{\rm apo} \rangle$, $\langle r_{\rm peri} \rangle$, maximum and minimum 
distance from the Galaxy's center achieved by a star during its orbit) and 
$\langle Z_{\rm max} \rangle$ (maximum distance from the Galaxy's plane achieved 
by a star during its orbit) are  $\langle r_{\rm apo} \rangle$  $\sim$ 10 kpc, 
$\langle r_{\rm peri} \rangle$ $\sim$ 5 kpc, and $\langle Z_{\rm max} \rangle$ $\sim$ 3 kpc,
respectively.\par

By using the fiducial sample of stars employed to determine the mean rotational 
velocity and dispersion for the MWTD-like stellar population (1 kpc $<|Z|<$ 2 kpc, 
$-$1.1 $<$[Fe/H]$<$ $-$0.6, and [${\alpha}$/Fe] $>$ +0.25), we found that the mean 
Galactocentric distance for this component is $R$ = 8.6 $\pm$ 0.2 kpc.\par

The ratio $\langle V_{\phi} \rangle / \sigma$ (where $\sigma$ is the total 
velocity dispersion) for stellar populations in the Galaxy's disk is typically larger 
than 1. In the case of the MWTD-like component, this parameter has a value of $\sim$ 2.5. 
Moreover, its rotational velocity shows a clear dependence on the vertical distance. 
Such properties unequivocally identify the MWTD as a disk component.

\section{Discussion}

\subsection{On the Thin-Disk Contamination}

The thin-disk component is not well-represented among the SDSS DR7 calibration stars 
\citep{carollo2010}, because these stars were selected at Galactic latitudes that 
would avoid thin-disk stars. However, some contamination may still be present in our 
data, which can be largely removed by selecting stars at $|Z|$ $>$ 1 kpc. Previous 
analyses have shown that a portion of the thin-disk stellar population might possess 
higher [$\alpha$/Fe] ratios (+0.1 $<$ [$\alpha$/Fe] $<$ +0.2), lower metallicities 
([Fe/H] $\sim$ $-$0.5), and located at larger Galactocentric distance than the low-$\alpha$
metal-rich thin disk (the so-called outer-disk stars, $R$ $>$  9 kpc; \cite{bovy2012b}). 
The $\alpha$-rich portion of this stellar disk can reach larger vertical distances 
than the $\alpha$-poor counterpart. To test if the $\alpha$-rich thin-disk stars can 
affect our results, we constructed a new Figure 1 (labeled as Figure 6) by selecting 
stars at Galactocentric distance of 7 kpc $<$ $R$ $<$ 9 kpc. Such a cut would exclude 
the outer thin-disk stars, as well as a fraction of the TD, MWTD, and halo stars. \par
Figure 6 shows that the two prograde groups identified in the ($L_{\rm Z}, L_{\rm 
P}$) diagram (panel a) are still present in this new sample, although the 
more-prograde cluster contains a lower number of stars than the sample represented 
in Figure 1. In the ($L_{\rm Z}$, [Fe/H]) plane (panel b), stars with $+0.1$ 
$<$ [$\alpha$/Fe] $<$ $+0.2$ (low-$\alpha$), are represented by a cyan color, while 
stars with [$\alpha$/Fe] $>$ $+0.22$ (high-$\alpha$) are shown in red. From inspection 
of this figure, the low-$\alpha$ prograde component, dominated by TD stars, is clearly 
separated from the high-$\alpha$ prograde component, dominated by the MWTD-like 
population. \par
It is important to remark that the above low-$\alpha$ and high-$\alpha$ separation 
reflects the range of [$\alpha$/Fe] ratios exhibited by the two clusters at [Fe/H] 
$>$ $-$1.2 in panel 1a and 1b of Figure 2, and that the high-$\alpha$ (as defined 
above) prograde stellar population also includes TD stars.  In panel 1c and 1d, the 
peak in $L_{\rm Z}$ associated with the less-prograde component is well-represented, 
while the peak related to the more-prograde component is somewhat flattened. This 
is mainly due to the fact that the cut in Galactocentric distance ($R <$ 9 kpc) also 
removes a significant number of  TD stars.\par
We applied the mixture modeling analysis to the sub-samples of stars obtained for 
different cuts of the vertical distance, $|Z|$, starting from the sample with -1.2 
$<$ [Fe/H] $<$ $-$0.6, $\alpha$-element ratios [$\alpha$/Fe] $>$ $+0.2$,  and  7 kpc 
$<$ $R$ $<$ 9 kpc (Sample C). The results are reported in Table 2. The kinematic 
parameters of the MWTD-like stellar population remain unchanged; it is the 
dominant component in all three distributions, with a mean angular momentum of 1100 
$-$ 1200 kpc km s$^{-1}$. The main differences between these results and those 
reported for Sample A in Table 1 are: (1) a lower mean vertical angular momentum 
for the TD close to the Galactic plane, $\langle L_{\rm Z} \rangle \sim$ 1700 
kpc km s$^{-1}$ ( $\sim$ 1900 kpc km s$^{-1}$ when 7 kpc $<$ R $<$ 10 kpc is used), 
and (2) the TD component is not present at vertical distances 2 kpc $< |Z| <$ 3 
kpc (the best fit was obtained with two components only). We suspect that by removing 
stars with $R >$ 9 kpc, the TD stellar population is significantly reduced, affecting 
the presence of such populations at these distances.\par

In previous works \citep{lee2011b, bovy2012b}, it was found that the outer 
thin disk is located in the metallicity range $-$ 0.7 $<$ [Fe/H] $< -$ 0.2,  with 
0.0 $<$ [$\alpha$/Fe] $<$ $+0.2$, and resides at Galactocentric distance $R >$ 9 kpc. 
In this abundance range, both the thin-disk and TD populations are present, and it 
is difficult to accurately quantify the thin-disk contamination, because this 
component is not well-represented in our sample. \par
Based on the results reported in previous works, we considered outer thin-disk stars 
to be those in the metallicity range $-$0.7 $<$ [Fe/H] $<$ $-$0.2 with $\alpha$-abundance 
ratios 0.0 $<$ [$\alpha$/Fe] $<$ +0.2, and located at $R >$ 9 kpc and 0 kpc $< |Z| 
<$ 2 kpc. The total number of stars located at $R >$ 9 kpc in our sample is N = 4527, 
while the number of likely outer thin-disk stars is N = 637, corresponding to a 
contamination of 14\%.
\clearpage

\subsection{Rapidly Rotating TD Stars}

In Figure 4, the peak values of $L_{\rm Z}$ for the sub-samples in different cuts 
of $|Z|$, obtained from Sample B, are larger ($\sim$ 2000 kpc km s$^{-1}$) than 
those expected for a distribution dominated by TD stars ($\sim$ 1600 kpc kms$^{-1}$, 
assuming the Sun's position at R = 8.5 kpc). Sample B is selected in a range of 
metallicity and $\alpha$-elements abundance ($-$0.9 $<$ [Fe/H] $<$ $-$0.6 and +0.1 $<$ [${\alpha}$/Fe] $<$ +0.2) that exclude the majority of thin-disk stars (see for example Figure 4 and Figure 2 in \cite{hayden2015} and \cite{cheng2012}, respectively). In this range of abundances, however, the outer thin-disk stars have been identified, $-$0.7 $<$ [Fe/H] 
$<$ $-$0.2, and 0.0 $<$ [$\alpha$/Fe] $<$ +0.2 \citep{lee2011b,bovy2012b}. The total number of stars in Sample B is 415, while the number of likely outer thin-disk stars, selected according to the above criteria, is 77. This means that Sample B has a contamination from “likely” outer thin-disk stars on the order of $\sim$ 18\%.. 
This would explain the large values of $L_{\rm Z}$ obtained for the TD. To investigate 
further, we explored the trend of the mean rotational velocity, 
$\langle$V$_{\phi}$$\rangle$, for Sample B ($-$0.9 $<$ [Fe/H] $<$ $-$0.6; +0.1 
$<$ [$\alpha$/Fe] $<$ +0.2) as a function of metallicity. It is known that thin-disk 
stars show a negative rotational velocity gradient ($\langle V_{\phi} \rangle / 
\Delta \langle |Z| \rangle$) as the metallicity increases, while TD stars exhibit 
a positive or zero velocity gradient \citep{lee2011b}. Figure 7 shows the mean 
Galactocentric rotational velocity, as a function of the metallicity, for Sample B, 
in two different intervals of vertical distance, 0 kpc $< |Z| <$ 1 kpc (black), 
and 1 kpc $< |Z| <$ 2 kpc (red). Visual inspection of this figure reveals that the 
mean rotational velocity has a zero or slightly positive gradient as the metallicity 
increases. Therefore, it is unlikely that Sample B has significant contamination from 
outer thin-disk stars. We conclude that, in the range of abundances where Sample B 
is selected, the majority of the stars belong to the TD, while other works claim that, 
in such an interval of metallicity and $\alpha$-abundance, the outer thin-disk stars are 
still present \citep{lee2011b,bovy2012b}. It remains unclear the reason 
why these TD stars possess such large rotational velocities, requiring further 
investigation.

\section{Implications for the formation of the MWTD}

How could a second stellar population in the TD have formed, and what are the 
implications for the assembly history of the Galaxy?\par
In a previous work \citep{carollo2010}, it was suggested that there may exist a 
connection between the MWTD-like component and the Monoceros Ring (an over-density 
of stars in a large area of the sky approximately parallel to the Galactic plane 
towards the anti-center of the Galaxy, in the Galactic latitude range 
10$^{\circ}$  $< |b| <$ 35$^{\circ}$, and spanning a large area in Galactic 
longitude, covering most of the second and third quadrants \citep{newberg2002,ibata2003}, on the basis of its mean metallicity, [Fe/H] $\sim$ $-$1.0 , 
and its rotational velocity, $\sim$ 110 km s$^{-1}$ \citep{ivezic2008}.  The origin 
of the Monoceros Ring has been a matter of debate, and recent evidence tends to support 
the hypothesis that it is not the debris of an accreted ancient satellite, but probably 
a part of the wobbly Galactic disk \citep{deboer2018,sheffield2018} 
induced by the tidal interaction with massive satellites such as the Sagittarius dwarf 
and/or the Large Magellanic Cloud \citep{laporte2018}. If this is the case, then the 
MWTD observed near the Sun may simply be a gravitationally heated/warped ensemble 
of TD stars, exhibiting a slower mean rotation than the TD, due to asymmetric drift.  
However, this simple scenario cannot explain the systematically lower metalliticites 
and higher [${\alpha}$/Fe] ratios of the MWTD stars compared with TD stars, as found 
in this work.\par
Instead, the MWTD stars might be more ancient TD stars, heated and 
raised to higher vertical distances, caused by a Gaia-Enceladus-like merger about 
10 billion years ago, in the course of the formation of the Milky Way's IH 
\citep{helmi2018,bignone2019}; the main body of the currently observed TD may be formed by the 
more recent merging of a dwarf satellite and associated disk heating 
\citep{gallart2019}. \par
Another possibility for explaining the MWTD stars is that they are originally formed 
at lower Galactocentric radii, $R \sim$ 5 kpc, than the Sun, before the main body of 
the TD formed, accounting for their lower $L_{\rm Z}$, lower [Fe/H], higher [${\alpha}$/Fe], and probably older ages than the TD. Then, internal dynamical process in this stellar 
disk, such as bar instability and/or spiral arm formation may have induced radial 
migration of these stars to the Solar Neighborhood.  Alternatively, the transport 
of stars might be caused by the merging of Gaia-Enceladus, inducing a non-axisymmetric 
structure such as a bar and/or spiral in this early disk, before the formation of 
the main body of the TD. \par
As an alternative scenario, the MWTD could result from partially circularized debris 
of a relatively massive accreted satellite.  The circularization occurs through 
dynamical friction, which favors a satellite of mass comparable to the SMC.  The 
observed metallicity of the MWTD would also be consistent with a satellite of such 
mass, through the mass-metallicity relation for dwarf galaxies \citep{kirby2013}.  
Deposition of satellite stars into the Galactic disk in this way would  suggest the 
possibility that dark matter from the satellite is also deposited into the disk. To 
assess the validity of these formation scenarios, it is worth further exploring 
state-of-the-art high-resolution numerical simulations for the formation of the 
Milky Way's disk and halo system in a consistent manner.\par

\acknowledgments 
The authors thank the anonymous referee for the valuable suggestions.
DC acknowledges support from the ESO Scientific Visitor Programme and support by Sonderforschungsbereich SFB 881 "The Milky Way System" (subproject A9) of the German Research Foundation (DFG). MC and MNI 
acknowledge partial support by MEXT Grant-in-Aid for Scientific
Research (No. 17H01101 and 18H04334 for MC, 18H05437 for MC and MNI). KCF acknowledges 
support from Australian Research Council grant DP160103747. TCB acknowledges partial 
support from grant PHY 14-30152; Physics Frontier Center/JINA Center for the 
Evolution of the Elements (JINA-CEE), awarded by the US National Science Foundation. 
YSL acknowledges support from the National Research Foundation (NRF) of Korea grant 
funded by the Ministry of Science and ICT (No.2017R1A5A1070354 and  
NRF-2018R1A2B6003961). PBT acknowledges partial support from Fondecyt 1150334 and UNAB Grant 10/19. CB 
acknowledges support by Sonderforschungsbereich SFB 881 "The Milky Way System" 
(subproject A9) of the German Research Foundation (DFG).

{}

\clearpage

\begin{deluxetable}{ccllc}
\tablewidth{0pt}
\tablenum{1}
\tablecaption{Parameters from Mixture Modeling Analysis for Sample A}
\tablehead{
\colhead{$|Z|$} &
\colhead{N} &
\colhead{$\langle L_{\rm Z} \rangle$} &
\colhead{$\sigma_{L_{\rm Z}}$} &
\colhead{P} \\
\colhead{(kpc)} &
\colhead{ } &
\colhead{(kpc km~s$^{-1})$} &
\colhead{(kpc km~s$^{-1})$} &
\colhead{\%} 
}
\startdata
0$-$1   &  TD  &  1904 $\pm$ 89  &  144 $\pm$ 133 & 43 \\
           &        &                 &                 &      \\
           &  MWTD  & 1267$\pm$72   &  353 $\pm$ 77 & 38 \\
           &        &                 &                 &      \\
          &   IH &    $-$8$\pm$199  & 443 $\pm$ 123 & 19   \\
          &        &                 &                 &       \\
1$-$2   &  TD  &  1744 $\pm$ 231  &  210 $\pm$ 34 & 26  \\
          &        &                 &                 &      \\
           &  MWTD  &  1173 $\pm$ 51  & 363 $\pm$ 31  & 60  \\
 &        &                 &                 &      \\
          &   IH &  134 $\pm$ 200  &  656 $\pm$ 85 & 14 \\
 &        &                 &                 &      \\
2$-$3   &   TD &  1774 $\pm$ 231  & 236 $\pm$ 111  &8  \\
 &        &                 &                 &      \\
           &   MWTD &  1076$\pm$ 96  &  416 $\pm$ 63 & 62 \\
 &        &                 &                 &      \\
          &    IH &  -198$\pm$179  &  473 $\pm$ 92 & 30 \\
 &        &                 &                 &      \\
\enddata
%\tablecomments{}
\end{deluxetable}

\clearpage

\clearpage
\begin{deluxetable}{ccllc}
\tablewidth{0pt}
\tablenum{2}
\tablecaption{Parameters from Mixture Modeling Analysis for Sample C}
\tablehead{
\colhead{$|Z|$} &
\colhead{N} &
\colhead{$\langle L_{\rm Z} \rangle$} &
\colhead{$\sigma_{L_{\rm Z}}$} &
\colhead{P} \\
\colhead{(kpc)} &
\colhead{ } &
\colhead{(kpc km~s$^{-1})$} &
\colhead{(kpc km~s$^{-1})$} &
\colhead{\%} 
}
\startdata
0$-$1   &  TD  &  1698 $\pm$ 166  &  200 $\pm$ 78 & 31 \\
           &        &                 &                 &      \\
           &  MWTD  & 1193 $\pm$ 93   & 157 $\pm$ 100 & 41 \\
           &        &                 &                 &      \\
          &   IH &  427 $\pm$ 224  & 490 $\pm$ 124 & 28   \\
          &        &                 &                 &       \\
1$-$2   &  TD  &  1691 $\pm$ 62  &  151 $\pm$ 49 & 10  \\
          &        &                 &                 &      \\
           &  MWTD  &  1176 $\pm$ 38  & 327 $\pm$ 25 & 74   \\
          &        &                 &                 &     \\
          &   IH &  134 $\pm$ 200  &  656 $\pm$ 85 & 15 \\
          &        &                 &                 &      \\
2$-$3   &       &                &                   &  \\
          &        &                 &                 &      \\
           &   MWTD &  1076 $\pm$ 96  &  416 $\pm$ 63 & 82 \\
          &        &                 &                 &      \\
          &    IH &  $-$198 $\pm$ 179  &  473 $\pm$ 92 & 18 \\
          &        &                 &                 &      \\
\enddata
%\tablecomments{}
\end{deluxetable}

\clearpage

\begin{figure}[!ht]
\begin{center}
\includegraphics[angle=90, scale = 0.6]{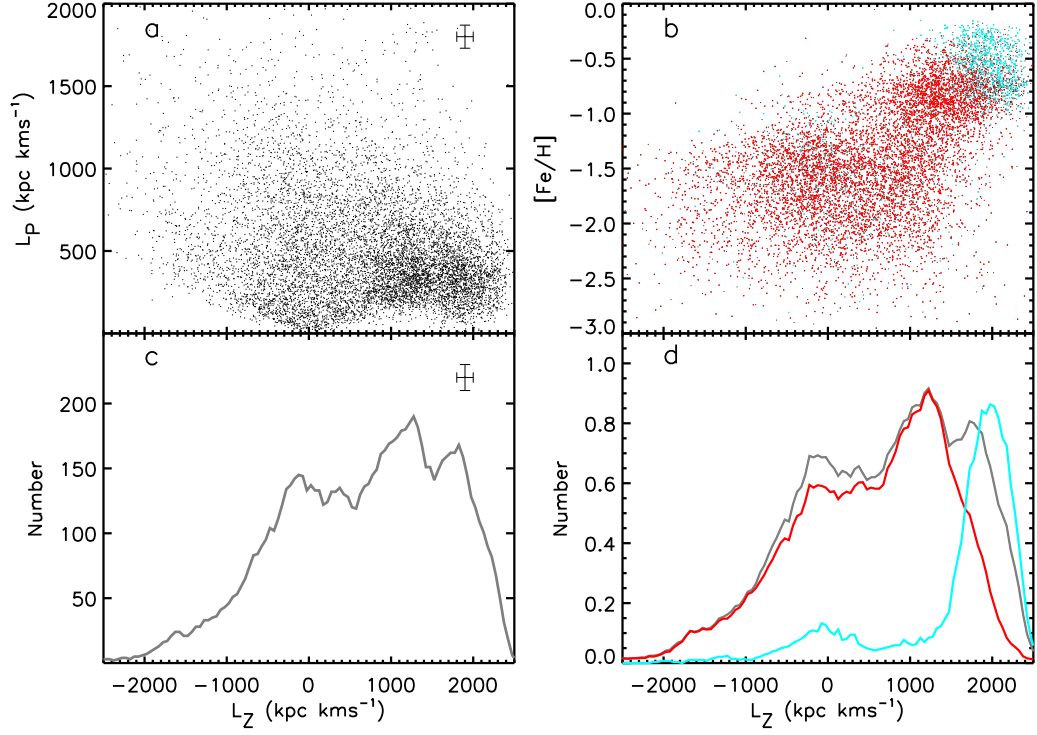}
\end{center}
\caption{The distribution of the entire sample of stars, obtained by matching the 
SDSS / SEGUE calibration stars with Gaia DR2, and by adopting the criteria described 
in the text. Panel (a) shows the distribution in the ($L_{\rm Z}, L_{\rm P}$) plane. 
Panel (b) shows the distribution in the ($L_{\rm Z}$, [Fe/H]) plane, color coded by 
low-$\alpha$ (cyan) and high-$\alpha$ (red). In the bottom panels, the density plots 
of $L_{\rm Z}$ are shown for the stars in panel a and b (panel c and d, respectively). 
In panel (d), the sample is color-coded as in panel (b).  The error bars in panel 1a show the average error in the angular momenta L$_{\rm Z}$ and L$_{\rm P}$, while in panel 1c denote the Poisson error for the two most significant peaks of the distribution.}
\end{figure}

\clearpage

\begin{figure}[!ht]
\begin{center}
%\hspace{2cm}
\includegraphics[angle=90, scale = 0.65]{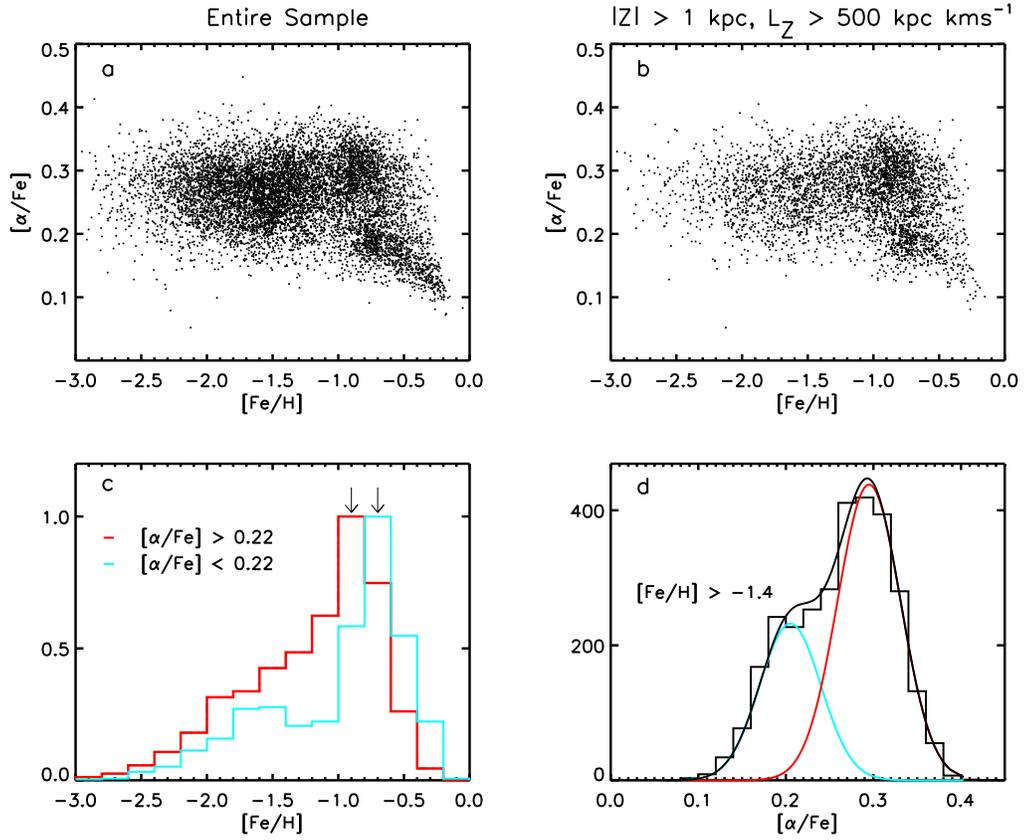}
\end{center}
\caption{Panel (a) shows the chemical abundance of the entire sample of stars, 
obtained by matching the SDSS / SEGUE calibration stars with Gaia DR2, and by adopting 
the criteria described in the text. In this chemical-abundance space three different 
components can be distinguished: the overlapping thin-disk and TD stellar 
populations, the MWTD-like component, and the halo system. Panel (b) represents the sample of stars 
located at larger distance from the Galactic plane ($|Z| >$ 1 kpc) and $L_{\rm Z} 
>$ 500 kpc km s$^{-1}$ (less thin-disk contamination and fewer IH stars). The two 
dominant groups, TD and MWTD-like, are clearly recognized, along with some halo stars. 
Panel (c) shows the metallicity distribution function for the sample in panel (b), 
but separated into two sub-samples according to their [$\alpha$/Fe] abundances. The two arrow symbols denote the location of the maxima of the distribution. In 
panel (d), the [$\alpha$/Fe] distribution is shown for the same sample, with an 
additional cut in metallicity to reduce halo contamination. The cyan and 
red lines are the individual Gaussian distributions included in the mixture model analysis applied to this last sample.} %\label{rv}
\end{figure}

\newpage

\begin{figure}[!ht]
%\begin{center}
\hspace{2cm}
\includegraphics[angle=0, scale = 0.7]{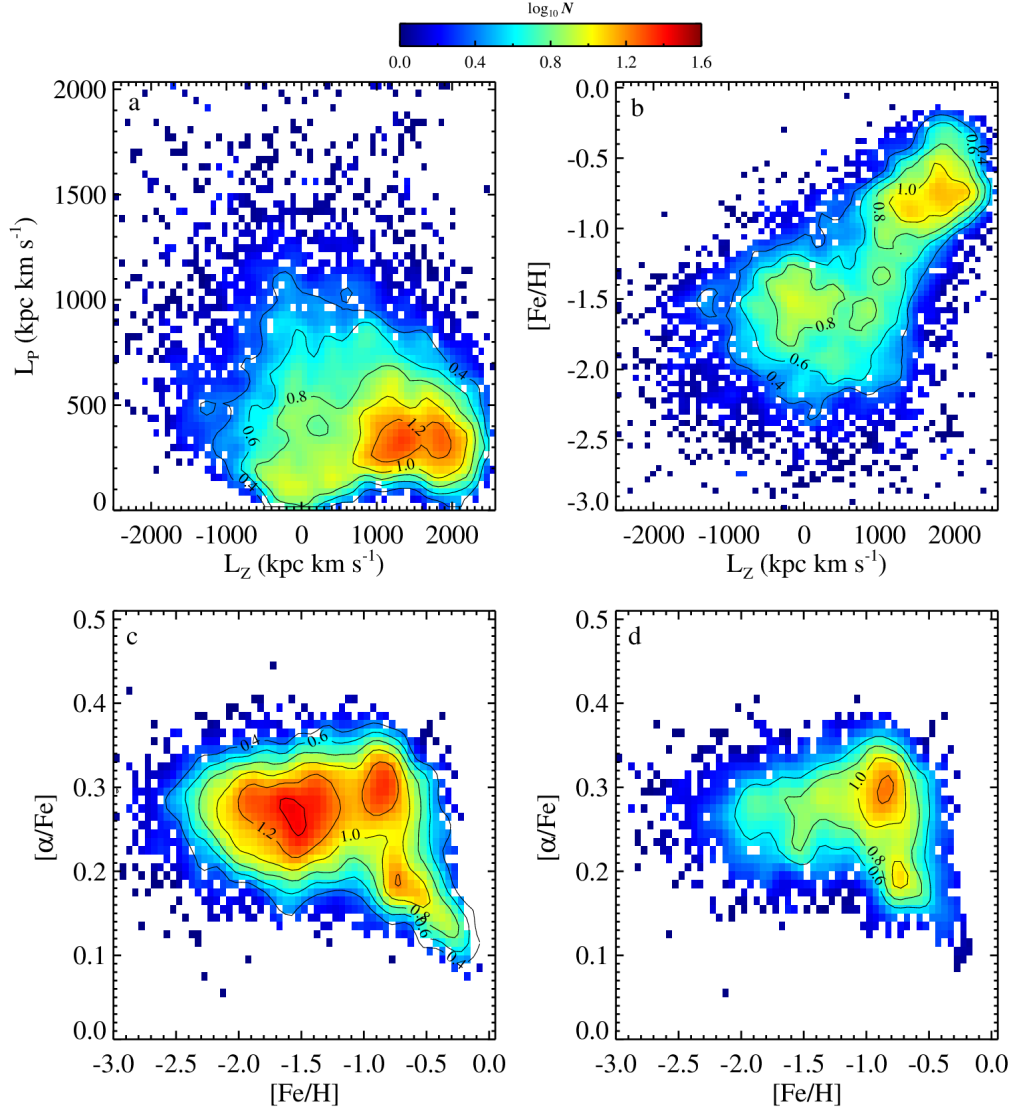}
%\end{center}
\vspace{0.5cm}
\caption{Distribution of the logarithmic number density in the ($L_{\rm Z}, L_{\rm 
P}$) and ([Fe/H], $L_{\rm Z}$) planes, for the entire sample (panels a and b), 
over-plotted with equidensity contours. The two prograde components ($L_{\rm Z} >$ 0 
kpc km s$^{-1}$) can be clearly identified with the high-density reddish clusters 
with equidensity contour of 1.2 (panel a) and 1.0 (panel b). In panel (a), each bin 
is 75 by 35 kpc km s$^{-1}$, while in panel (b) the bins are 0.04 dex by 75 kpc km s$^{-1}$. 
Panel (c) represents the distribution of the logarithmic number density in the 
([$\alpha$/Fe], [Fe/H]) plane for the entire sample, while panel d shows the 
distribution for the sample obtained by performing the selection $|Z| >$ 1 kpc and 
$L_{\rm Z} >$ 500 kpc km s$^{-1}$. In panels (c) and (d), the bin size is ([$\alpha$/Fe], 
[Fe/H]) = (0.05 dex, 0.01 dex). All the density distributions are smoothed with a 
Gaussian kernel of 3 pixels. } 
\end{figure}

\begin{figure}[!ht]
%\begin{center}
\hspace{2.5cm}
\includegraphics[angle=0, scale = 0.65]{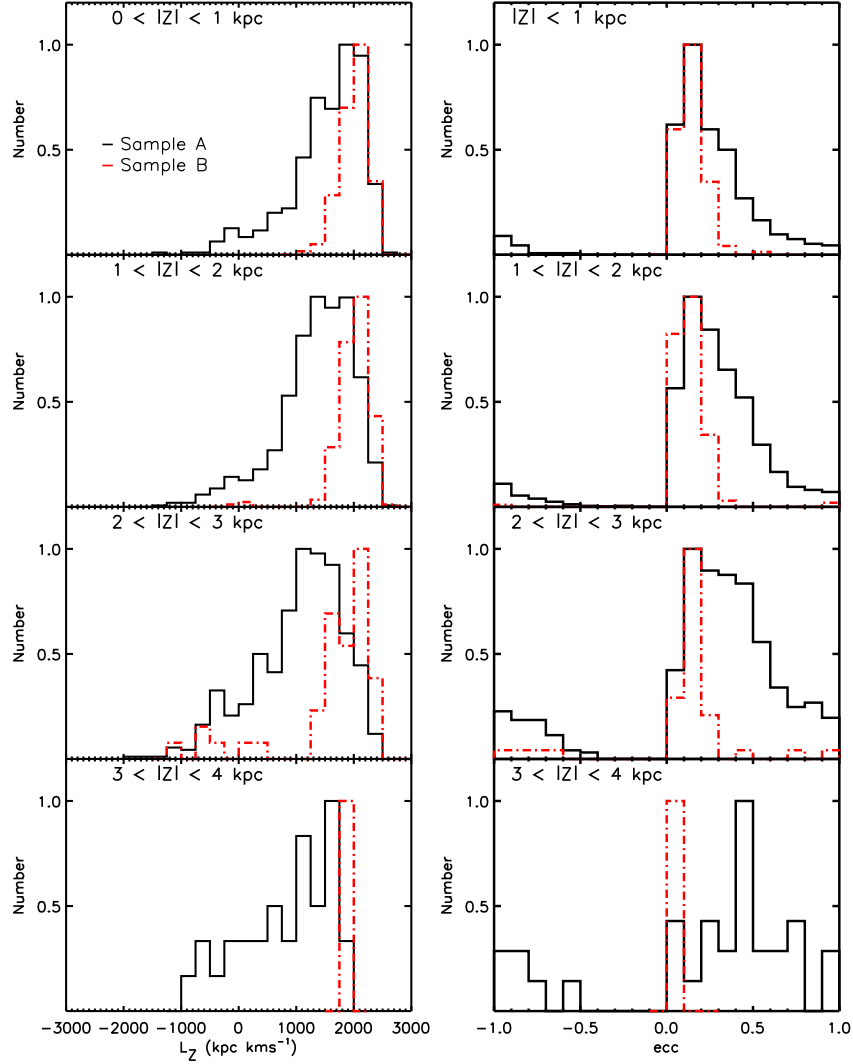}
%\end{center}
\caption{The left column of panels shows the distribution of $L_{\rm Z}$ for different 
cuts in vertical distance for samples A ($-$1.2 $<$ [Fe/H] $<$ $-$0.6 and 
[$\alpha$/Fe] $>$ +0.2, black continuous line) and B ($-$0.9 $<$ [Fe/H] $<$ $-$0.6 
and +0.1 $<$ [$\alpha$/Fe] $<$ +0.2, red dot-dashed line), respectively, as described 
in the text. The right column of panels represent the orbital eccentricity distribution for these samples. Note that the eccentricity is 'signed', in order to differentiate stars with retrograde (negative ecc) from those with prograde (positive ecc) orbits. The distribution are normilized to one to allow comparison between the samples.} 
\end{figure}

\clearpage

\begin{figure}[!ht]
%\begin{center}
\hspace{3cm}
\includegraphics[angle=0, scale = 0.55]{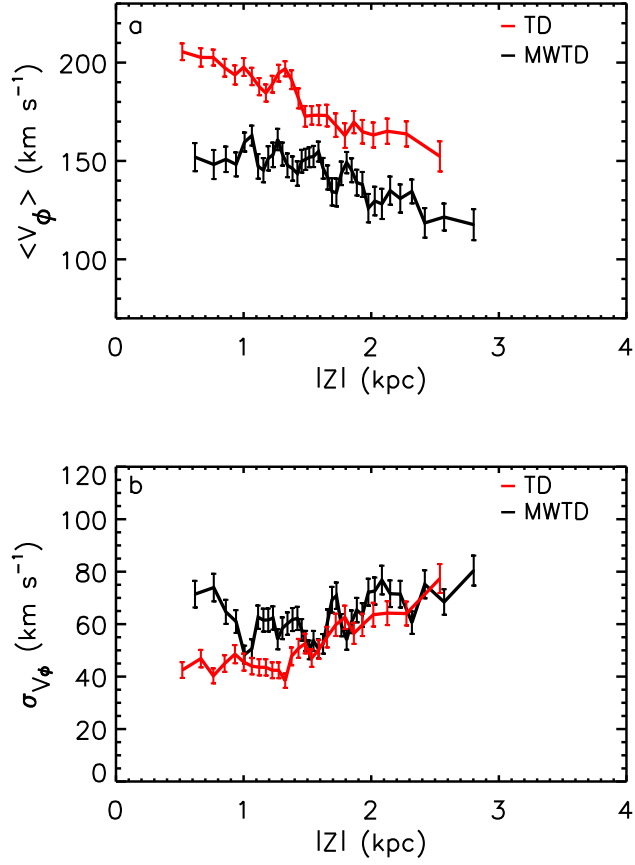}
%\end{center}
\vspace{0.5cm}
\caption{Mean Galactocentric rotational velocity (panel a) and dispersion (panel b), 
as a function of the vertical distance from the Galaxy's plane, $|Z|$, for the TD 
(red) and MWTD (black). Each bin represents 100 stars with an overlap of 70 stars 
per bin. } 
\end{figure}

\begin{figure}[!ht]
\begin{center}
%\hspace{4cm}
\includegraphics[angle=90, scale = 0.6]{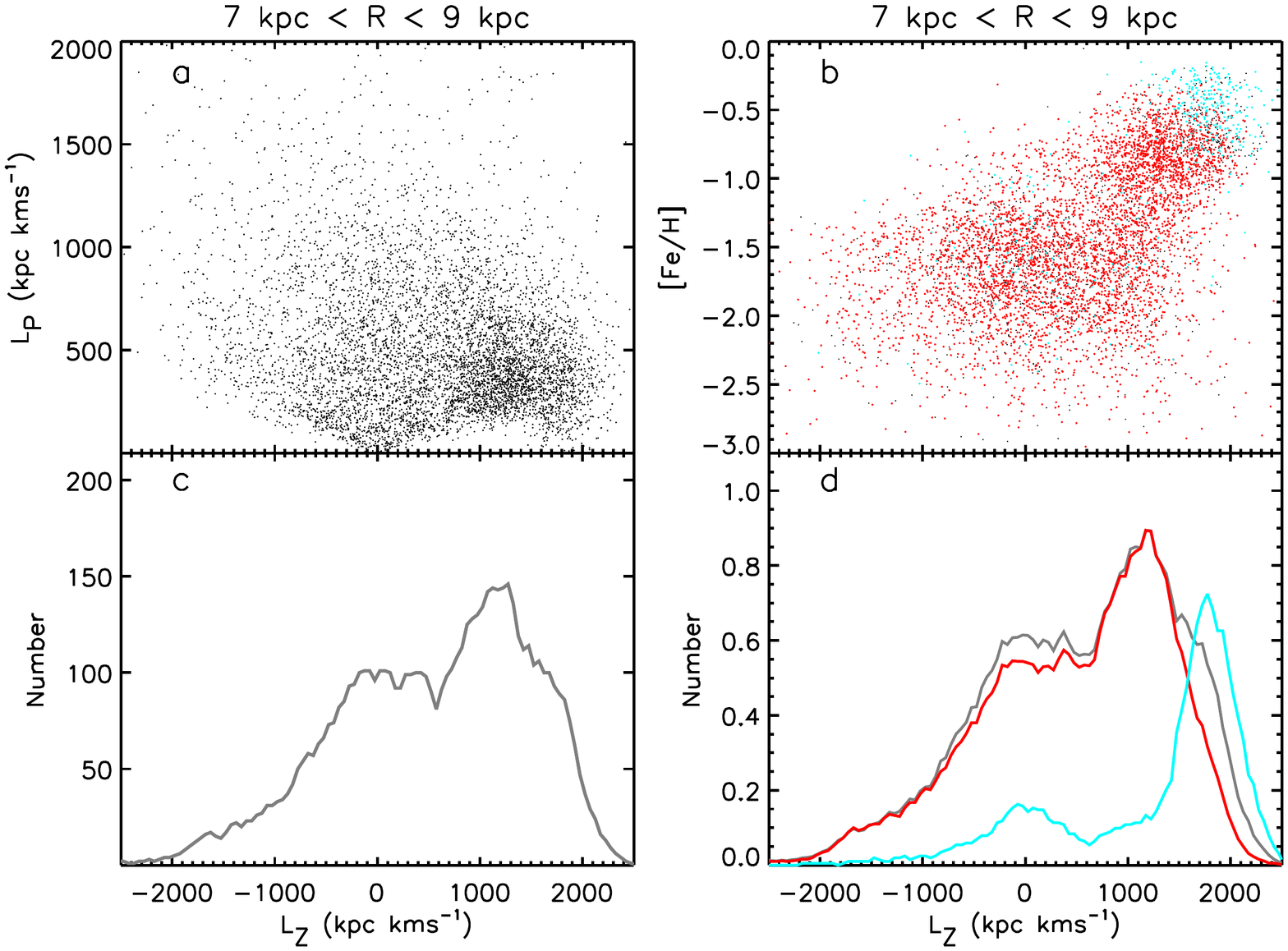}
\end{center}
\vspace{0.5cm}
\caption{The same as Figure 1, but for a sub-sample of stars selected by removing 
the stars at $R >$ 9 kpc  from the entire sample (7 kpc $<$ $R$ $<$ 10 kpc,  N = 9258 
stars), to avoid contamination from outer thin-disk stars.  In the right hand panels, 
the color coding indicates  low-$\alpha$ (cyan) and high-$\alpha$ (red) stars.} 
\end{figure}

\clearpage
\begin{figure}[!ht]
%\begin{center}
\hspace{3.5cm}
\includegraphics[angle=0, scale = 0.6]{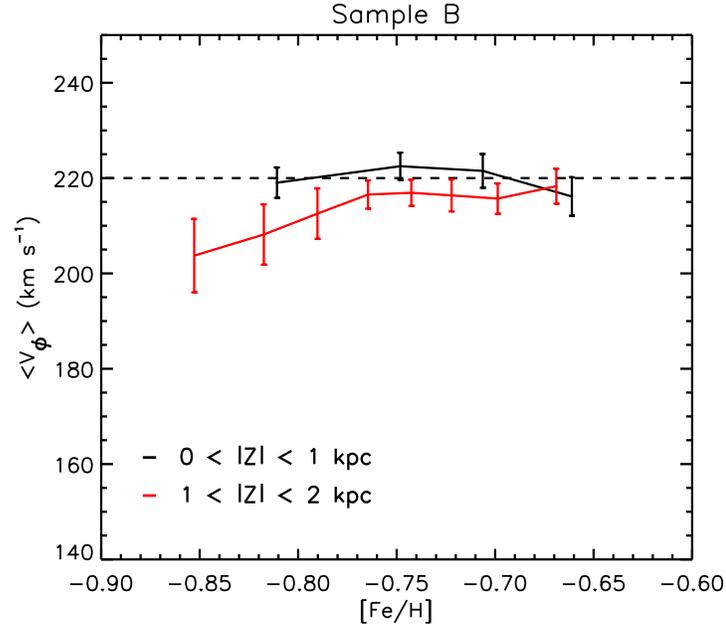}
%\end{center}
\vspace{-0.2cm}
\caption{Mean Galactocentric rotational velocity as a function of the metallicity 
for Sample B ($-$0.9 $<$ [Fe/H] $<$ $-$0.6; +0.1 $<$ [$\alpha$/Fe] $<$ +0.2). The 
black curve shows the trend for stars in the vertical distance range  0 kpc $< |Z| 
<$ 1 kpc, while the red sequence includes stars at larger distances from the Galactic 
plane (1 kpc $< |Z| <$ 2 kpc). Each bin of metallicity contains 50 stars. The dashed 
line shows the value of the LSR circular velocity.
} 
\end{figure}

\clearpage

\appendix

\begin{center}
{A1. Effects of Biases on Parallaxes}
\end{center}

There are two potential problems to take into account when using Gaia DR2 parallaxes: the issues that could arise by calculating the distance as the inverse of parallax, and the well-known parallax zero-point offset.\par
In this analysis the stellar distances were estimated by taking the inverse of parallax for stars with relative error $\sigma_{\pi}/\pi <$ 0.2. However, distance determination from parallax measurements is an inference problem and was extensively discussed by \citet{bailer-jones2018}. The top-right panel of Figure A1 shows the comparison between the distances obtained by taking the inverse of parallax (d$_{\rm g}$) with those derived by \citet{bailer-jones2018} (d$_{\rm BJ}$) for the sample of stars satisfying the condition $\sigma_{\pi}/\pi <$ 0.2. As can be appreciated from this figure, there is very good agreement up to $\sim$ 2 kpc, then d$_{\rm g}$ tends to be somewhat over-estimated with respect to d$_{\rm BJ}$, with deviations on the order of $\sim$ (0.2 - 0.6) kpc that increases with distance. To assess if such deviations may influence the results presented in this paper, the kinematic parameters are evaluated by adopting the Bailer-Jones distances. The ($L_{\rm Z}, L_{\rm P}$) diagram obtained with these distances is shown in the bottom-left panel of Figure A1. This new angular momenta diagram is identical to the top-left panel of Figure 1, and exhibit two groups of stars with prograde rotation at $L_{\rm Z}$ $\sim$ 1800 kpc km s$^{-1}$ and $L_{\rm Z}$ $\sim$ 1200 kpc km s$^{-1}$. This excercise shows that the distance obtained by the inverse of parallax for the sample of stars with $\sigma_{\pi}/\pi <$ 0.2 is a robust determination.\par

\citet{lindegren2018} showed that parallax measurements are affected by an average zero-point offset of $\delta_{\pi}$ = $-$0.029 mas. This bias appears to be dependent on multiple variables, such as stellar magnitudes, colors, sky position, and the employed stellar sample, and it affects primarily distant stars (small parallaxes). Recent works report a slightly higher value of the parallax zero-point offset, on the order of $\delta_{\pi}$ = $-$0.05 mas \citep{graczyk2019, leung2019, schonrich2019,zinn2019}. In order to establish the important of this offset on the results, we performed the bulk of the analysis by adding a constant value of 0.054 mas to the parallaxes. The assumption of a costant parallax zero-point offset can be justified by following the results reported in \citet{leung2019}, where it is shown that the constant offset model is in very good agreement with the multi-variate offset model up to 10 kpc for a sample of $\sim$ 265,000 stars in common between APOGEE DR14 \citep{holtzman2015,garciaperez2016,abolfathi2018} and Gaia DR2. \par 

The top-right panel of Figure A1 shows the distance obtained with the inverse of parallax corrected for the bias, as a function of the distance inferred with the Bailer-Jones's method. The agreement between the two distances is improved, and the maximum deviation is only 0.2 kpc. Again, to assess the importance of the parallax correction for the zero-point offset on the results reported in this paper, we derived the kinematic parameters with these new distances. The bottom-right panel of Figure A1 shows the ($L_{\rm Z}, L_{\rm P}$) diagram for the sample of stars with  relative error on parallax $<$ 20\%, which now contains a larger number of stars with respect to the sample obtained with no zero-point offset. This can be explained by the fact that many stars are closer and are included in the local volume of 4 kpc. The bottom-right panel clearly shows that the two prograde groups of stars at $L_{\rm Z}$ $\sim$ 1800 kpc km s$^{-1}$ and $L_{\rm Z}$ $\sim$ 1200 kpc km s$^{-1}$ are now even better defined than before, thanks also to the increased number of stars in the local volume, below 4 kpc. \par

The vertical angular momentum distribution for the samples A and B obtained with the parallax zero-point offset correction is shown in Figure A2. A comparison with Figure 4 shows that, at 0 kpc $< |Z| <$ 3 kpc, both $L_{\rm Z}$ and the eccentricity distribution for the two samples are almost identical to those reported in the three top panels of this figure. Farther from the Galactic plane, at 3 kpc $< |Z| <$ 4 kpc, there are only a few stars due to the slightly shorter distances obtained with the addition of the parallax zero-point offset. The results of the mixture model analysis applied to these new distributions are reported in Table A1. A comparison with Table 1 shows that the values of mean angular momentum and dispersion for the TD, MWTD, and IH are very similar. Close to the Galactic plane (0 kpc $< |Z| <$ 1 kpc), the distribution is dominated by the TD and MWTD-like components, while at larger vertical distances the MWTD-like component is dominant, with mean angular momentum of $\sim$ 1100 to $\sim$ 1200  kpc km s$^{-1}$. The inner-halo stellar population accounts for 15\%, 17\%, and 25\% of the total number of stars in Sample A, respectively, while the TD-dominated component contributes 45\%, 40\%, and 18\%, at 0 kpc $< |$Z$| <$ 1 kpc, 1 kpc $< |$Z$| <$ 2 kpc, and 2 kpc $< |$Z$| <$ 3 kpc, respectively.\par 

The mean Galactocentric rotational velocity and dispersion for the MWTD-like stellar population was obtained by adopting the fiducial sample of stars described in Section 3 (1 kpc $<$ $|Z|$ $<$ 2 kpc, $-$1.1 $<$ [Fe/H] $<$ $-$0.6, and 
[${\alpha}$/Fe] $>$ +0.25). The results are  $\langle V_{\phi} \rangle$ = 150 $\pm$ 2 
km s$^{-1}$ and $\sigma_{V_{\phi}}$ = 58 $\pm$ 2 km s$^{-1}$ which are totally in agreement with the values reported in Section 3, obtained without adding the zero-point offset bias to the parallaxes.

\begin{center}
{A2. An Additional Sample\\}
\end{center}

An additional sample of stars, selected by cross-matching the SDSS DR15 \citep{aguado2019} 
with Gaia DR2, was  employed to obtain a separate confirmation of the MWTD component 
in phase space and  chemical space. The sample contains only stars with available spectroscopy, and for which the radial velocities and the chemical abundances are 
determined through the SSPP pipeline. The [$\alpha$/Fe] ratios were taken from SDSS 
DR7 \citep{Abazajian2009}.\par

We selected stars with a relative error on parallax  $\sigma_{\pi}/{\pi} <$ 0.2, 
and, calculated distances using the inverse of parallax, as for the calibration-star 
sample (with no parallax zero-point offset correction). The resulting number of stars is $\sim$ 68,000, and the majority have 
heliocentric distances d $<$ 4 kpc. Kinematical parameters are obtained by assuming 
the Sun's location at 8.5 kpc \citep{ghez2008} from the Galaxy's center and 
correction for Sun's motion of $(U,V,W)$ =  ($-$9, 12, 7) km s$^{-1}$. 
The circular velocity at the position of the Sun is 220 km s$^{-1}$ \citep{bovy2012a}. 
Orbital angular momenta are computed by using the procedures described above.\par

Figure A3 shows the distribution of the selected stars in the ($L_{\rm Z}, L_{\rm 
P}$) plane.  It is worth noting that in this additional sample that the thin-disk component 
is better represented than in the calibration-star sample used in our main analysis.  
Therefore, to reduce the contamination from this stellar population, we 
only plot stars with metallicity [Fe/H] $<$ $-$0.6 and $\alpha$-element abundance ratio 
[$\alpha$/Fe] $>$ +0.1. We also included an additional cut in metallicity ([Fe/H] 
$>$ $-$1.6) to reduce the contamination from halo stars. In panel (a), the more-prograde 
(TD-dominated) and less-prograde components are clearly recognizable as the two 
clusters, with angular momentum $L_{\rm Z} \sim$ 1800 kpc km s$^{-1}$ and $L_{\rm Z} 
\sim$ 1200 kpc km s$^{-1}$, that we associate with the TD-dominated and the MWTD-like 
stellar populations, respectively. The density distribution shown in panel (b) exhibits 
two peaks in $L_{\rm Z}$, as in the calibration-star sample (Figure 1), confirming the detection of the MWTD-like component in SDSS DR15 as well.\\

\setcounter{table}{0}
\renewcommand{\thetable}{A\arabic{table}}
\begin{deluxetable}{ccllc}
\tablewidth{0pt}
%\tablenum{1}
\tablecaption{Parameters from Mixture Modeling Analysis for Sample A (With Parallax Zero-Point Correction)}
\tablehead{
\colhead{$|Z|$} &
\colhead{N} &
\colhead{$\langle L_{\rm Z} \rangle$} &
\colhead{$\sigma_{L_{\rm Z}}$} &
\colhead{P} \\
\colhead{(kpc)} &
\colhead{ } &
\colhead{(kpc km~s$^{-1})$} &
\colhead{(kpc km~s$^{-1})$} &
\colhead{\%} 
}
\startdata
0$-$1   &  TD  &  1917 $\pm$ 35  &  187 $\pm$ 19 & 45 \\
           &        &                 &                 &      \\
           &  MWTD  & 1318$\pm$56   &  251 $\pm$ 61 &40 \\
           &        &                 &                 &      \\
          &   IH &    372$\pm$226  & 550 $\pm$ 114 & 15   \\
          &        &                 &                 &       \\
1$-$2   &  TD  &  1819 $\pm$ 40  &  217 $\pm$ 19 & 30  \\
          &        &                 &                 &      \\
           &  MWTD  &  1210 $\pm$ 51  & 320 $\pm$ 46  & 53  \\
 &        &                 &                 &      \\
          &   IH &  184 $\pm$ 150  &  546 $\pm$ 70 & 17 \\
 &        &                 &                 &      \\
2$-$3   &   TD &  1696 $\pm$ 124  & 230 $\pm$ 60  &18  \\
 &        &                 &                 &      \\
           &   MWTD &  1108$\pm$ 125  &  359 $\pm$ 89 & 57 \\
 &        &                 &                 &      \\
          &    IH &  -2.5$\pm$198  &  465 $\pm$ 100 & 25 \\
 &        &                 &                 &      \\
\enddata
%\tablecomments{}
\end{deluxetable}

\clearpage

\renewcommand{\thefigure}{A\arabic{figure}}
\setcounter{figure}{0}

\begin{figure}[!ht]
%\begin{center}
\hspace{2cm}
\includegraphics[angle=90, scale = 0.6]{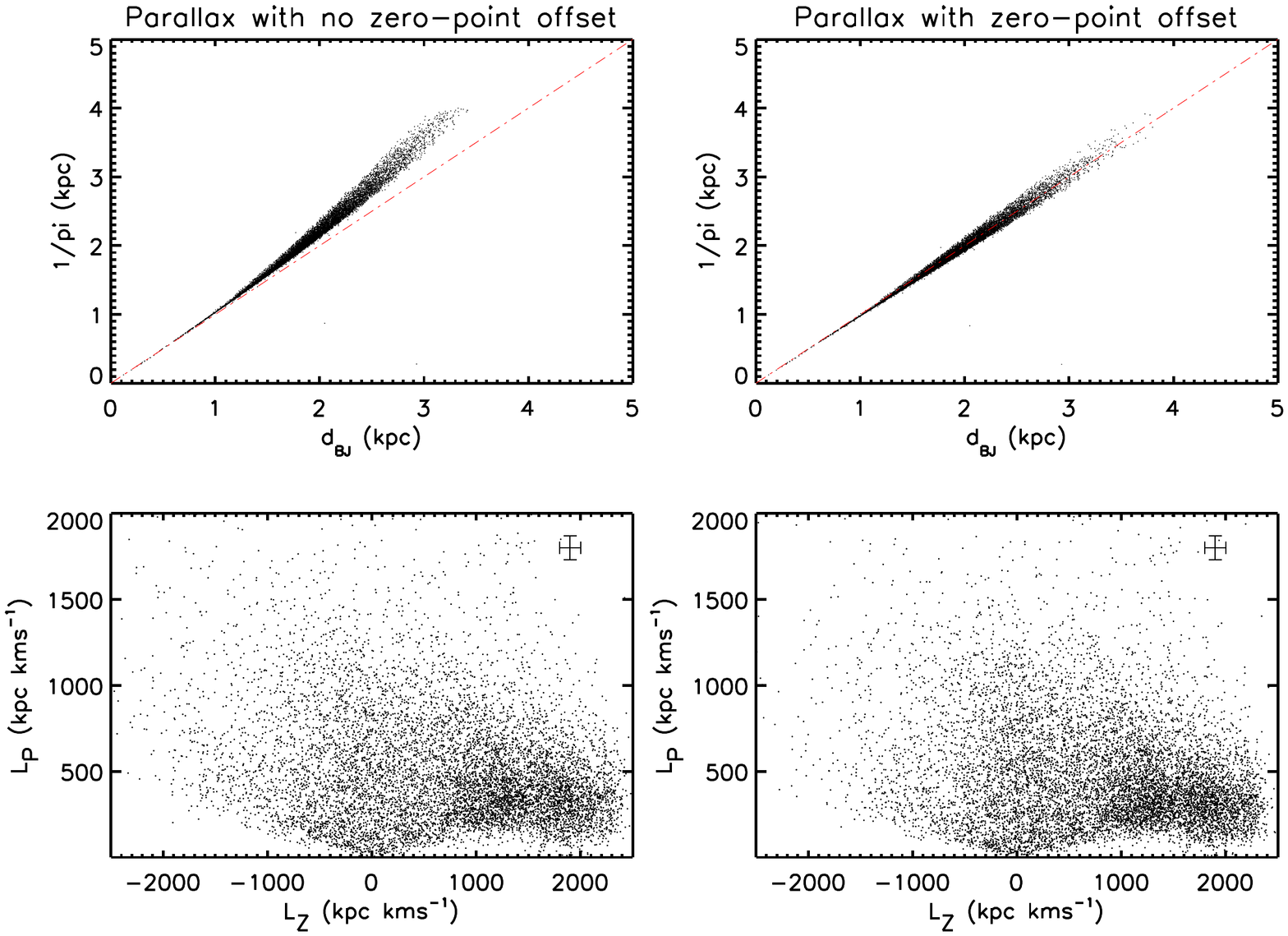}
%\end{center}
\vspace{0.5cm}
\caption{Top-left panel: SDSS stellar distances obtained by using the inverse of parallax with no zero-point offset vs those derived by \citet{bailer-jones2018}. Top-right panel: SDSS stellar distances obtained by using the inverse of parallax with zero-point offset added vs those derived by \citet{bailer-jones2018}. Bottom-left panel: the distribution in the ($L_{\rm Z}, L_{\rm P}$) plane for the entire sample of stars and kinematics obtained by adopting the distances derived by \citet{bailer-jones2018}. 
 Bottom-right: the distribution in the ($L_{\rm Z}, L_{\rm P}$) plane for the entire sample of stars and kinematics obtained by adopting the distances derived by the inverse of parallax with the zero-point offset added.} 
%\label{figa1}
\end{figure}

\renewcommand{\thefigure}{A\arabic{figure}}
\setcounter{figure}{1}

\begin{figure}[!ht]
%\begin{center}
\hspace{3cm}
\includegraphics[angle=0, scale = 0.65]{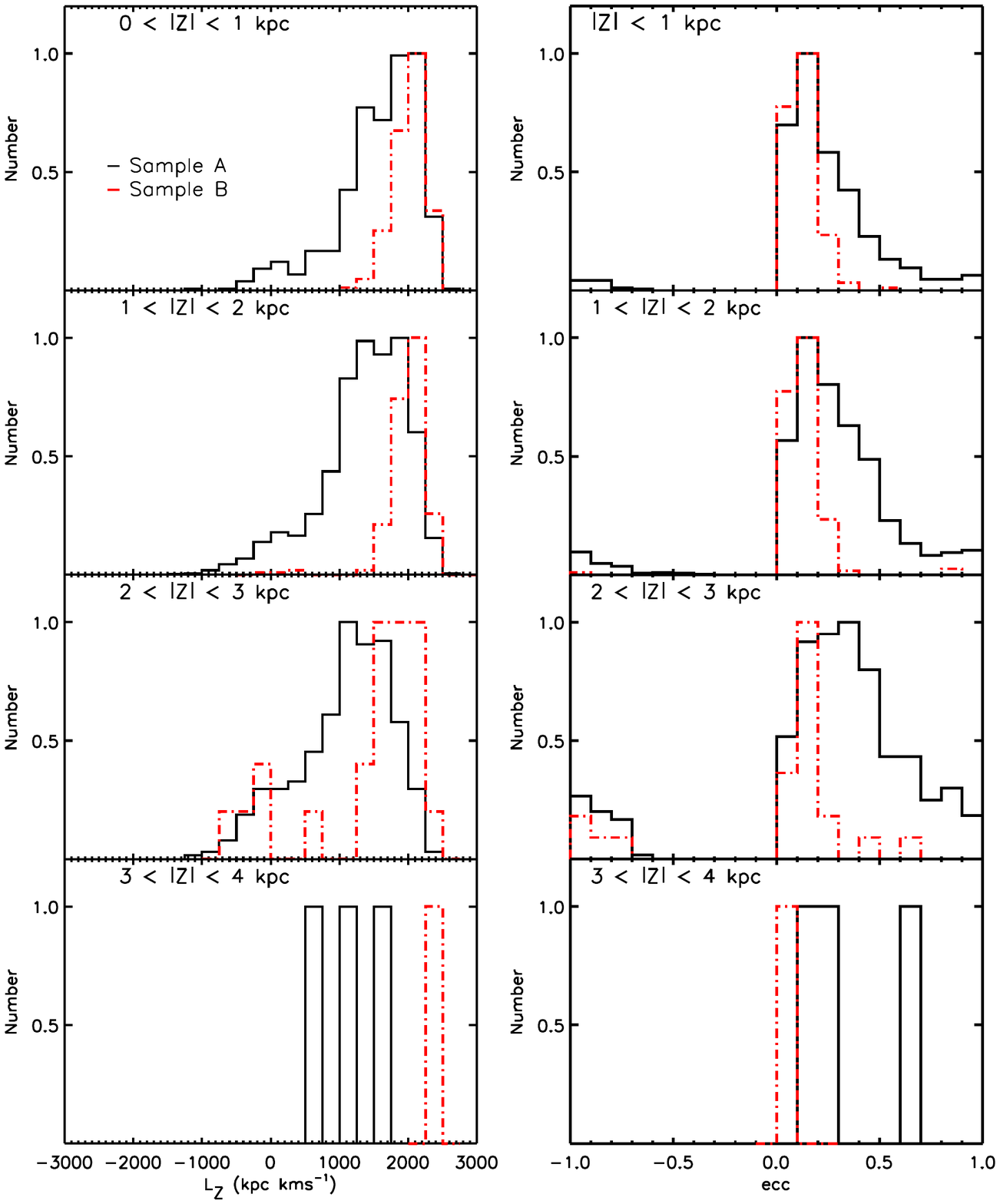}
%\end{center}
\vspace{0.5cm}
\caption{The same as in Figure 4, but with angular momentum and eccentricity obtained by using the inverse of parallax with the zero-point offset added.} 
%\label{figa1}
\end{figure}

\renewcommand{\thefigure}{A\arabic{figure}}
\setcounter{figure}{2}

\begin{figure}[!ht]
%\begin{center}
\hspace{4cm}
\includegraphics[angle=0, scale = 0.6]{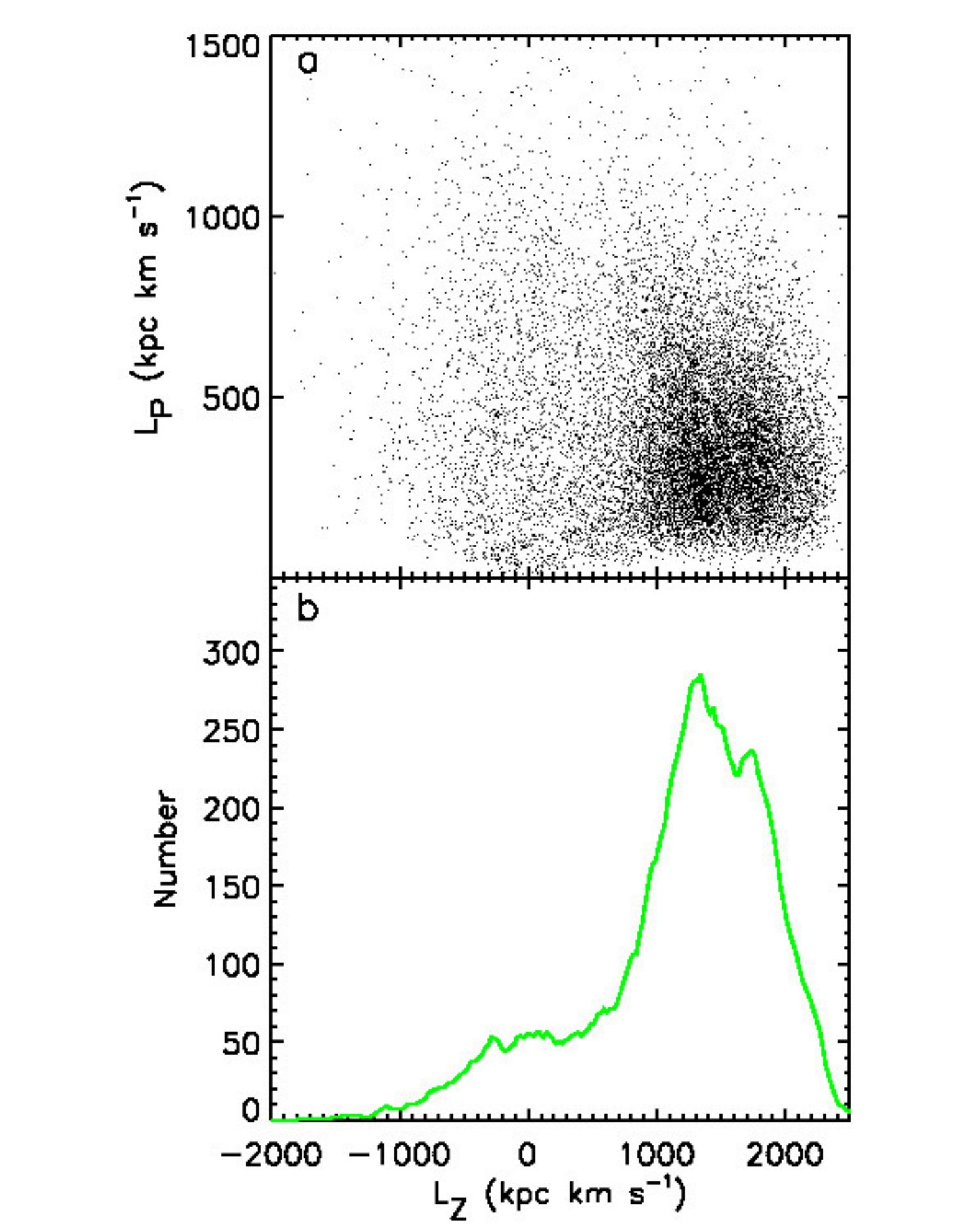}
%\end{center}
\vspace{0.5cm}
\caption{The distribution of the SDSS DR15 stars, cross-matched with Gaia DR2, in 
the ($L_{\rm Z}, L_{\rm P}$) plane (with no parallax zero-point correction). Panel (a) represents the sample obtained by 
applying the cuts in metallicity and $\alpha$-elements described in the text. Panel (b) 
is a density plot of $L_{\rm Z}$.} 
%\label{figa1}
\end{figure}


\begin{thebibliography}{}
\bibitem[Abazajian et al. (2009)]{Abazajian2009} Abazajian, K.N., Adelman-McCarthy, J.K., Agueros, MA., et al. , 2009, \apjs, 182, 543
\bibitem[Abolfathi et al. (2018)]{abolfathi2018} Abolfathi B., et al., 2018, \apjs, 235, 42
\bibitem[Aguado et al. (2019)]{aguado2019} Aguado, D.S., Ahumada, Romina, Almeida, A., et al. 2019, 240, 23 
\bibitem[Allende Prieto et al. (2007)]{Allende2007} Allende Prieto, C., Beers, T.C., Sivarani, T., et~al., 2007, \aj, 136, 2070
\bibitem[Bailer-Jones et al. (2018)]{bailer-jones2018} Bailer-Jones, C.A.L., Rybizki, J., Fouesneau, M., et al., 2018, \apj, 156,58
\bibitem[Beers et al. (2006)]{beers2006} Beers, T. C., Lee, Y. S., Sivarani, T., et al., 2006, Mem. S.A.It.,77,1171
\bibitem[Beers et al. (2012)]{beers2012} Beers, T. C., Carollo, D., Ivezic, Z., et al., 2012, \apj, 746, 34
\bibitem[Beers et al. (2014)]{beers2014} Beers, T.C., Norris, J.E., Placco, V.M., et al., 2014, \apj, 794, 58
\bibitem[Bignone et al. (2019)]{bignone2019} Bignone, L.A., Helmi, A., Tissera, P.B, 2019, arXiv:1908.07080
\bibitem[Bovy et al. (2012a)]{bovy2012a} Bovy, J., Allende Prieto, C., Beers, T.C., et al. , 2012, \apj, 759, 131
\bibitem[Bovy et al. (2012b)]{bovy2012b} Bovy, J. Rix, H.-W., Liu, C., et al., 2012, \apj, 753, 148
\bibitem[Brook et al. (2007)]{brook2007} Brook, C. ~B., Kawata, D., Martel, H., Gibson, B. K., \& Scannapeico, E., 2007 EAS Publications Series, 24, 269
\bibitem[Brown et al. (2018)]{brown2018} Brown, A.,  et al., 2018, \aa, 616, A1
\bibitem[Bullock et al. (2005)]{bullock2005} Bullock, J. S., \& Johnston, K. V., 2005, \apj, 635, 931
\bibitem[Carollo et al. (2007)]{carollo2007} Carollo, D., Beers, T.~C., Lee, Y.~S., et~al., 2007, \nat, 450, 1020
\bibitem[Carollo et al. (2010)]{carollo2010} Carollo, D., Beers, T.~C., Chiba, M., et al. 2010, \apj, 712, 692
\bibitem[Castelli \& Kurucz (2003)]{castelli2003} Castelli, F. \& Kurucz, R. L. , 2003, Modelling of Stellar Atmospheres. 210th IAU Symposium, Edited by N. Piskunov, W.W. Weiss, and D.F. Gray, 210, A20 
\bibitem[Cheng et al. (2012)]{cheng2012} Cheng, J.Y., Rockosi C.M., Morrison, F., et al., 2012, \apj, 752, 51
\bibitem[Chiba \& Beers (2000)]{chiba2000} Chiba, M. \& Beers, T.C., 2000, \aj, 119, 2843
\bibitem[Cui et al. (2012)]{cui2012} Cui, X.-Q., Zhao, Y.-H., Chu, Y.-Q., et al. 2012, RAA, 12, 1197
\bibitem[de Boer et al. (2018)]{deboer2018} de Boer, T.J.L., Belokurov, V. \& Koposov, S.E., 2018, \mnras, 473, 647
\bibitem[Hayden et al. (2015)]{hayden2015} Hayden, M.R., Bovy, J., Holtzman, J.A., et al. , 2015. \apj, 808, 132 
\bibitem[Haywood et al. (2013)]{haywood2013}	Haywood, M., Di Matteo, P., Lehnert, M.D., Katz, D., Gomez, A., 2013, \apj, 560, A109
\bibitem[Holtzman et al. (2015)]{holtzman2015} Holtzman J. A., et al., 2015, \aj, 150, 148
\bibitem[Garcia Perez et al. (2016)]{garciaperez2016} Garcia Perez A. E., et al., 2016, \aj, 151, 144
\bibitem[Gallart et al. (2019)]{gallart2019} Gallart, C., Bernard, E.J., Brook, C.B., et al., 2019, \nat, arXiv: 1901:02900  
\bibitem[Ghez et al. (2008)]{ghez2008} Ghez, A. M., Salim, S., Weinberg, N.N., et al.,  2008, \apj, 689, 1044
\bibitem[Gilmore, G., \& Reid, N. (1983)]{gilmore1983} Gilmore, G., \& Reid, N., 1983, \mnras, 202, 1025
\bibitem[Graczyk et al. (2019)]{graczyk2019} Graczyk, D., Pietrzynski, G., \& Gieren, W., 2019, \apj, 872, 1
\bibitem[Grillmair \& Carlin (2016)]{grillmair2016} Grillmair, C.J. \& Carlin, J.L., 2016, Springer International Publishing Switzerland, Switzerland, 420,87
\bibitem[Helmi et al. (2018)]{helmi2018}	Helmi, A., Babusiaux, C. Koppelman, H.H., et al., 2018, \nat, 563, 85 
\bibitem[Lee et al. (2006)]{lee2006} Lee, Y. S., Beers, T.C., Sivarani, T., et al. , 2006, BAAS, 209, 168.15
\bibitem[Luri et al. (2018)]{luri2018} Luri, X., et al., 2018, \aa, 616, 9L
\bibitem[Ibata et al. (2003)]{ibata2003} Ibata R.,Irwin M. J., Lewis G. F., Ferguson A. M. N., Tanvir N., 2003, \mnras, 340, 21L
\bibitem[Ivezic et al. (2008)]{ivezic2008}	Ivezic, Z., Sesar, B., Juric, M., et al. , 2008, \apj, 684, 287
\bibitem[Laporte et al. (2018)]{laporte2018} Laporte, C.F.P., Johnston, K.V., \& Tzanidakis, A., 2018, \mnras, 1427, 1436
\bibitem[Lee et al. (2008a)]{lee2008a}  Lee, Y. S., Beers, T.C., Sivarani, T., et al., 2008, \aj, 136, 2022
\bibitem[Lee et al. (2008b)]{lee2008b}  Lee, Y. S., Beers, T.C., Sivarani, T., et al., 2008, \aj, 136, 2050
\bibitem[Lee et al. (2011a)]{lee2011a} Lee, Y.~S., Beers, T.~C., Allende Prieto, C., et al., 2011, \apj, 141, 90 
\bibitem[Lee et al. (2011b)]{lee2011b} Lee, Y.~S., Beers, T.~C., An, D., et al., 2011, \apj, 738, 187 
\bibitem[Leung \& Bovy (2019)]{leung2019} Leung H. W., Bovy J., 2019, \mnras, 489, 2079 
\bibitem[Lindegren et al. (2018)]{lindegren2018} Lindegren, L., Hernandez, J., Bombrun, A., et al., 2018, \aa, 616, A2
\bibitem[Mihalas \& Binney  (1981)]{mihalas1981} Mihalas, D., \& Binney, J. , 1981, Galactic Astronomy (San Francisco: Freeman)  
\bibitem[Myeong et al. (2019)]{myeong2019} Myeong, G.C., Vasiliev, E., Iorio, G., et al., 2019, \mnras, 488, 1235 
\bibitem[Moore et al. (2006)]{moore2006} Moore, B., Diemand, J., Madau, P., Zemp, M., \& Stadel, J., 2006, \mnras, 368, 563
\bibitem[Morrison et al. (1990)]{morrison1990} Morrison, H.L, Flynn, C., \& Freeman, K. , 1990, \aj, 100, 1191
\bibitem[Newberg et al. (2002)]{newberg2002} Newberg, H.J., Yanny, B., Rockosi, C., et al. , 2002, \apj, 569, 245
\bibitem[Prugniel \& Soubiran (2001)]{prugniel2001} Prugniel, Ph. \& Soubiran, C.A.  2001, \aa, 369, 1048
\bibitem[Recio-Blanco et al. (2014)]{recioblanco2014} Recio-Blanco, A., de Laverny, P., Kordopatis, G., et al., 2014, \aa, 567, A5
\bibitem[Rix \& Bovy (2013)]{rix2013} Rix, H. W. \& Bovy, J. , 2013, \aa, 21, 61
\bibitem[Robin et al. (2014)]{robin2014} Robin, A. C., Reylé, C., Fliri, J., et al. 2014, \aa, 569, A13
\bibitem[Robin et al. (2017)]{robin2017} Robin, A.C., Bienayme, O., Fernandez-Trincado, J.G., Reyle, C., 2017, \aa, 605, A1
\bibitem[Savino \& Posti  (2019)]{savino2019} Savino, A. \& Posti, L. , 2019, \aa, 624, L9
\bibitem[Schonrich et al. (2019)]{schonrich2019} Schonrich, R., McMillan, P., \& Eyer, L., 2019, 487, 3568
\bibitem[Sheffield et al. (2018)]{sheffield2018} Sheffield, A.A., Johnston, K.V., Price-Whelan, A.M., et al. 2018, \apj, 854, 1
\bibitem[Schlegel D. J., Finkbeiner, D. P., \& Davis (1998)]{schlegel1998} Schlegel, D. J., Finkbeiner, D. P., \& Davis, 1998, \apj, 500, 525
\bibitem[Tian et al. (2019)]{tian2019} Tian, H., Liu, C., Xu, Y, Xue, X., 2019, \apj, 871, 184 
\bibitem[Tissera et al. (2010)]{tissera2010} Tissera, P.B., White, S.D.M., Pedrosa, S., Scannapieco, C. , 2010, \mnras, 406, 922
\bibitem[Kirby et al. (2013)]{kirby2013} Kirby, E. N., Cohen, J.G., Guhathakurta, P., et al., 2013, \apj, 779, 102
\bibitem[Yanny et al. (2009)]{yanny2009} Yanny, B., Rockosi, C., Newberg, H.J., et al., 2009, \aj, 137, 4377
\bibitem[York et al. (2000)]{york2000} York, D. G. et al., 2000, \aj, 120, 1579
\bibitem[Yoshii (1982)]{yoshii1982} Yoshii,Y., 1982, PASJ, 34, 365
\bibitem[Zinn et al. (2019)]{zinn2019} Zinn, J.C., Pinsonneault, M.H., Huber, D., \& Dennis Stello, 2019, \apj, 878, 2
\bibliographystyle{apj}
\end{thebibliography}
\end{document}